\documentclass[a4paper]{article}
\usepackage{bm}
\usepackage{amsmath,amssymb}
\usepackage{url}
\usepackage{caption}
\usepackage[dvipdfmx]{graphicx}
\usepackage{arydshln}
\usepackage{mathtools}
\usepackage[top=3.0cm, bottom=3.5cm, left=2.5cm, right=2.5cm]{geometry}
\usepackage[colorlinks=true, linkcolor=blue, urlcolor=blue]{hyperref}
\usepackage{colortbl}

\begin{document}
\baselineskip=1.25\baselineskip


\begin{center}
{\Large
\textbf{Bayesian predictive probability based on a bivariate index}
}\\\vspace{4pt}
{\Large
\textbf{vector for single-arm phase II study with binary efficacy and}
}\\\vspace{4pt}
{\Large
\textbf{safety endpoints}
}
\end{center}

\

\begin{center}
Takuya Yoshimoto$^{1}$, Satoru Shinoda$^{2}$, Kouji Yamamoto$^{2}$ and Kouji Tahata$^{3}$

\

$^{1}${Biometrics Department, Chugai Pharmaceutical Co., Ltd., Chuo-ku, Tokyo, 103-8324, Japan}\\
$^{2}${Department of Biostatistics, Yokohama City University School of Medicine, Yokohama City, Kanagawa, 236-0004, Japan}\\
$^{3}${Department of Information Sciences, Faculty of Science and Technology, Tokyo University of Science, Noda City, Chiba, 278-8510, Japan}

\

E-mail: \href{mailto:yoshimoto.takuya61@chugai-pharm.co.jp}{\nolinkurl{yoshimoto.takuya61@chugai-pharm.co.jp}}

\

Last update: July 31, 2023
\end{center}
\
\begin{abstract}
In oncology, phase II studies are crucial for clinical development plans as such studies identify potent agents with sufficient activity to continue development in the subsequent phase III trials. Traditionally, phase II studies are single-arm studies, with the primary endpoint being short-term treatment efficacy. However, drug safety is also an important consideration. In the context of such multiple-outcome designs, predictive probability-based Bayesian monitoring strategies have been developed to assess whether a clinical trial will provide enough evidence to continue with a phase III study at the scheduled end of the trial. Herein, we propose a new simple index vector for summarizing results that cannot be captured by existing strategies. Specifically, for each interim monitoring time point, we calculate the Bayesian predictive probability using our new index and use it to assign a go/no-go decision. 
Finally, simulation studies are performed to evaluate the operating characteristics of the design. The obtained results demonstrate that the proposed method makes appropriate interim go/no-go decisions.

\end{abstract}

\noindent
$\textbf{Keywords}$: Bayesian monitoring, efficacy and toxicity endpoints, index vector, predictive probability, single-arm phase II study.

\hypertarget{S1}{}
\section{Introduction}
In oncology, the primary objective of phase II clinical trial is commonly to evaluate the preliminary efficacy of a new treatment to identify agents with sufficient activity to continue development in the subsequent phase III studies. Accordingly, phase II studies play a key role throughout the clinical development plan. Typically, phase II oncology studies follow a single-arm study, with the primary endpoint being short-term treatment efficacy. Specifically, objective response based on the Response Evaluation Criteria in Solid Tumors (RECIST) \hyperlink{R9}{(Eisenhauer et al., 2009)} guidelines is commonly used as a primary endpoint in terms of the treatment efficacy. These studies routinely implement futility monitoring and interim go/no-go decision making to protect participants against ineffective treatments and to accelerate the clinical development. Numerous frequentist and Bayesian designs have been proposed for single-arm phase II studies. From the perspective of single-outcome designs, the well-known frequentist design is Simon's two-stage design \hyperlink{R26}{(Simon, 1989)}, which minimizes the maximum sample size or the expected sample size under the null hypothesis that the treatment is ineffective, while also controlling for type I and II error rates. For the Bayesian framework, \hyperlink{R29}{Thall and Simon (1994a)} provided practical guidelines on implementing an adaptive single-arm phase II study, which involved updating the Bayesian posterior probability after observing the data. \hyperlink{R16}{Lee and Liu (2008)} developed predictive probability-based Bayesian monitoring for a single-arm phase II clinical trial. They evaluated the probability of achieving the desired outcome at the planned end of the trial, conditional on accumulated interim data. In addition, \hyperlink{R23}{Sambucini (2021b)} extended the designs of \hyperlink{R16}{Lee and Liu (2008)} by considering the uncertainty of the historical control. Furthermore, \hyperlink{R36}{Yin et al. (2012)} used the predictive probability to construct a randomized phase II design based on Bayesian adaptive randomization.
A predictive distribution approach considers decisions upon final analysis based on data that will be available in the future, whereas a posterior distribution approach relies only on data at intermediate time points. Definitive conclusions are often difficult to obtain from data at interim time points; therefore, the effectiveness of the final analysis demands that consideration be given to data that will be available in the future. As such, predictive distributions are easy to interpret as decision-making at the time of the final analysis. \hyperlink{R24}{Saville et al. (2014)} and \hyperlink{R3}{Berry (2004)} have summarized the advantages of using predictive distributions, which have been used in several clinical trials, designed using predictive probability (see, e.g., \hyperlink{R6}{Buzdar et al., 2005}; \hyperlink{R8}{Domchek et al., 2020}; \hyperlink{R18}{Murai et al., 2021}; \hyperlink{R28}{Tap et al., 2012)}. 

Although the primary endpoint is commonly set as an efficacy outcome, situations may arise in which the safety outcome is of the same importance as the efficacy outcome.
From the perspective of a multiple-outcome design, several methods to bivariate binary outcomes that represent efficacy and toxicity have been presented. For example, \hyperlink{R5}{Bryant and Day (1995)} introduced a two-stage design with efficacy and toxicity instead of a single-efficacy endpoint. \hyperlink{R31}{Thall et al. (1995)} extended the designs of \hyperlink{R29}{Thall and Simon (1994a)} from single to multiple outcomes and accounted for the distinction between adverse and desirable outcomes. \hyperlink{R4}{Brutti et al. (2011)} also presented a Bayesian posterior probability-based approach that imposed a restrictive definition of the overall goodness of the therapy by controlling the number of responders who simultaneously do not experience adverse toxicity. Similarly, \hyperlink{R21}{Sumbucini (2019)} proposed a Bayesian decision-making method based on predictive probability, involving both efficacy and safety, with binary outcomes. 
Specifically, the decision-making was based on the marginal probability of experiencing toxicity and on the joint probability of simultaneously experiencing a response to the therapy without experiencing toxicity. Other studies have also focused on both efficacy and safety with binary outcomes (\hyperlink{R37}{Zhou et al., 2017}; \hyperlink{R12}{Guo et al., 2020}; \hyperlink{R22}{Sambucini, 2021a}). 
From the regulatory aspect of drug development, a guideline on Complex Innovative Trial Design has been issued by the United States Food and Drug Administration (FDA) (\hyperlink{R33}{Guidance for industry, 2020}), and the use of Bayesian is mentioned in the guideline. Recently, \hyperlink{R20}{Ruberg et al. (2023)} also stated that the widespread adoption of Bayesian approaches has the potential to be the most impactful tool for improving the development of new medicines.
Additionally, with regard to searching for an optimal dose, \hyperlink{R25}{Shah et al. (2021)} pointed out that for targeted drug and biological therapies, dose selection should be informed by exposure--response, efficacy, and safety data from early trials, rather than by automatically selecting the maximum tolerated dose (see also the FDA \hyperlink{R34}{Draft Guidance for Industry, 2023}).
Therefore, it is expected that Bayesian statistics will increasingly become used in clinical trials in the near future.

In this paper, we employ Bayesian single-arm designs with binary efficacy and safety outcomes that are based on predictive probability, an approach that is consistent with that prescribed by \hyperlink{R4}{Brutti et al. (2011)} and \hyperlink{R21}{Sumbucini (2019)}. Specifically, for the purpose of exemplification, we use an artificial dataset, as shown in Table \hyperlink{T1}{1}, with a small sample size. 

\begin{center}
\lbrack Tables 1(a) and (b) about here.\rbrack
\end{center}

\noindent
Table \hyperlink{T1}{1}(a) is more promising than Table \hyperlink{T1}{1}(b), as the joint probability of simultaneously experiencing efficacy and toxicity is substantially different between the tables. However, because of  the same marginal probability of experiencing toxicity and the joint probability of a responder who simultaneously does not experience adverse toxicity, \hyperlink{R4}{Brutti et al. (2011)} and \hyperlink{R21}{Sumbucini (2019)} could not capture these differences. Accordingly, we consider the probability of simultaneously being a non-responder to the therapy while experiencing toxicity. In this case, the natural idea is to consider the joint distribution of these probabilities; however, the elicitation of such a distribution is complicated because the marginal probability of toxicity and the probability of simultaneously being a non-responder to the therapy while experiencing toxicity are not mutually exclusive. In addition, computational burden is a well-known issue when calculating predictive probability (\hyperlink{R24}{Saville et al., 2014}); thus, constructing a simple index to summarize the results is desirable. Therefore, we consider an approach for decision-making by assuming the worst situation for the potential effect of a treatment and measuring the deviation from that situation. Given that the worst situation can be expressed as a particular probability distribution, it is natural to consider an index based on the difference (or similarity) between two probability distributions. Various divergences represented by the Kullback--Leibler Divergence (KLD) have been proposed for two distributions of a discrete random variable. In this paper, we focus on the Jensen--Shannon Divergence (JSD) (\hyperlink{R17}{Lin, 1991}) to address the specific situation whereby the KLD cannot generally be defined, and we propose a method to adaptively evaluate using a new index based on the JSD. The JSD has been used in many fields and has been applied in the design of clinical trials (see, e.g., \hyperlink{R10}{Fujikawa et al., 2020}). As efficacy and toxicity may not be independent, a method that accounts for their joint distribution would be appropriate in this situation. Therefore, we do not consider marginal models that lead to a decision that is independently obtained through the conjunction of two marginal distributions. In addition, we consider adding uncertainty to the historical control data using a prior distribution to the proposed methodology.

The remainder of this article is organized as follows: in Section \hyperlink{S2}{2}, we introduce the notation and propose a bivariate index to summarize data from clinical trials with binary outcomes. Here, we describe an asymptomatic approach, in addition to the Monte Carlo approach, to calculate the proposed predictive probability. In addition, we present a decision-making framework based on predictive probability using the proposed index.
A detailed illustration of the proposed procedure for each approach is presented in Section \hyperlink{S3}{3}. Section \hyperlink{S4}{4} describes the properties of the asymptotic approach using different hypothetical situations. In Section \hyperlink{S5}{5}, we present the results of simulation studies aimed at evaluating the operating characteristics of the proposed methodology. Finally, Section \hyperlink{S6}{6} presents a discussion along with the design parameter optimization.
All the computations presented in this paper were performed using R programming (\hyperlink{R19}{R Core Team 2023}).

\hypertarget{S2}{}
\section{Methods}
\hypertarget{S21}{}
\subsection{New bivariate index vector}

We consider a single-arm phase II study based on binary endpoints that represent the efficacy and toxicity data that are summarized in the $2\times 2$ square contingency tables in Table \hyperlink{T2}{2}, with the corresponding cell probabilities ${\bm p}=(p_{11}, p_{12}, p_{21}, p_{22})^{\prime}$ such that $\sum_{i}\sum_{j}p_{ij}=1$, where $\prime$ denotes the transpose. 

\begin{center}
\lbrack Table 2 about here.\rbrack
\end{center}

\noindent
Next, we consider the worst situation for the potential effect of a treatment by controlling the number of non-responders who simultaneously experience adverse toxicity. To examine this structure, along with the marginal probability of toxicity, we consider utility that is defined as the ratio of $p_{12}$ to $p_{21}$. Thereafter, we regard this utility as an index of efficacy. 
Next, we consider an approach where a decision on the potential effect of a drug is made by measuring the distance from a probability distribution. 
Let ${\bm q}=(q_{11}, q_{12}, q_{21}, q_{22})^{\prime}$ be discrete finite bivariate probability distribution that indicates the worst situation as well as $\bm p$. Furthermore, we denote $p^{*}_{ij}=p_{ij}/\left(p_{ij}+p_{ji}\right)$ and $q^{*}_{ij}=q_{ij}/\left(q_{ij}+q_{ji}\right)$ for $i=1,2$; $j=1,2$, respectively.
From the perspective of efficacy, we define the worst case as that expressed as a structure with $q^{*}_{12}=0$, assuming that efficacy does not depend on a diagonal cell. In the case of cytotoxic agents where efficacy and toxicity occur simultaneously, it may be necessary to make a decision based on the risk--benefit from $p_{11}$. However, given a situation where efficacy and safety are co-primary, it is natural to assume that the risk--benefit do not depend on a diagonal cell.
From the perspective of toxicity, we define the worst case as that expressed as a structure with $q_{\cdot 1}=1$. Then, we consider measuring the deviation from these situations. 
Because the worst situation can be expressed as a particular probability distribution, it is natural to consider an index based on the difference (or similarity) between two probability distributions. Then the KLD is used as it is the standard approach for calculating the difference between probability distributions. The KLD between $\bm p$ and $\bm q$ is shown as follows:
$$
{\rm KLD}({\bm p}:{\bm q})=\sum_{i}\sum_{j}p_{ij}\log\frac{p_{ij}}{q_{ij}}.
$$
Notably, ${\rm KLD}({\bm p}:{\bm q})$ is undefined if $q_{ij}=0$ and $p_{ij}\not=0$ for at least one pair of $i,j$. This means that the distribution $\bm p$ must be absolutely continuous with respect to the distribution $\bm q$ for ${\rm KLD}({\bm p}:{\bm q})$ to be defined (\hyperlink{R17}{Lin, 1991}).
However, we need to consider that the probability distribution $\bm q$ includes $q^{*}_{12}=0$ and $q_{\cdot 1}=1$ (then $q_{\cdot 2}=0$) from the perspectives of efficacy and toxicity, respectively. 
As such, the KLD cannot generally be defined in this situation; thus, to address this issue, we focus on the JSD as a measure of the difference between two distributions. ${\rm JSD}({\bm p}:{\bm q})$ is a well-grounded symmetrization of the KLD defined as
$$
{\rm JSD}({\bm p}:{\bm q})=\frac{1}{2}\sum_{i}\sum_{j}\left(p_{ij}\log\frac{2p_{ij}}{p_{ij}+q_{ij}}+q_{ij}\log\frac{2q_{ij}}{p_{ij}+q_{ij}}\right).
$$
A notable feature of ${\rm JSD}({\bm p}:{\bm q})$ is that this divergence requires that $\bm p$ and $\bm q$ be absolutely continuous with respect to each other. In addition, the JSD is always upper bounded by $\log 2$. Using this divergence, we propose new indexes for summarizing efficacy $\Phi_{eff}$ and toxicity $\Phi_{tox}$ as follows:
\begin{equation*}
\begin{split}
\displaystyle \Phi_{eff}&=\displaystyle\frac{1}{2\log 2}\left(p^{*}_{12}\log 2 + p^{*}_{21}\log\frac{2p^{*}_{21}}{p^{*}_{21}+1}+\log\frac{2}{p^{*}_{21}+1}\right),\\[3pt]
\displaystyle \Phi_{tox}&=\displaystyle\frac{1}{2\log 2}\left(p_{\cdot 2}\log 2 + p_{\cdot1}\log\frac{2p_{\cdot1}}{p_{\cdot1}+1}+\log\frac{2}{p_{\cdot1}+1}\right).
\end{split}
\end{equation*}
For details, see Appendix \hyperlink{A1}{A}. The index $\Phi_{eff}$: (i) lies between $0$ and $1$; (ii) $\Phi_{eff}=0$ if and only if the conditional probability $p^{*}_{12}=0$; and (iii) $\Phi_{eff}=1$ if and only if $p^{*}_{12}=1$. Similarly, the index $\Phi_{tox}$: (i) lies between $0$ and $1$; (ii) $\Phi_{tox}=0$ if and only if the marginal probability $p_{\cdot 1}=1$; and (iii) $\Phi_{tox}=1$ if and only if $p_{\cdot 1}=0$. This measure expresses the degree of departure from the most adverse case toward the most promising case in terms of efficacy and toxicity. Considering the joint association structure of $\Phi_{eff}$ and $\Phi_{tox}$, we propose the following bivariate index vector:
\hypertarget{E1}{}
\begin{equation}
{\bm \Phi} =\left(\Phi_{eff}, \Phi_{tox}\right)^{\prime},
\end{equation}
where ${\bm \Phi}$ is a $2\times 1$ vector. To derive the posterior distribution of ${\bm \Phi}$, an approach based on the Monte Carlo method (see, e.g., \hyperlink{R27}{Spiegelhalter, 2004}, p.103) is possible; however, this causes a substantial computational burden due to the complexity of the predictive distribution (see, Section \hyperlink{S22}{2.2}). Thus, we derive an asymptotic distribution for the posterior distribution of the proposed index vector. To estimate the indexes, $\hat{\bm{\Phi}}$ is given by $\bm {\Phi}$ with $\Phi_{eff}$ and $\Phi_{tox}$ replaced by $\hat{\Phi}_{eff}$ and $\hat{\Phi}_{tox}$ using $\hat{\bm{p}}$ as a consistent estimator of $\bm{p}$. With a sample size of $N$, the posterior distribution of $\sqrt{N}{(\hat{\bm \Phi}} - {\bm \Phi})$ has an asymptotically a bivariate normal distribution with a mean of zero and the following covariance matrix:
\hypertarget{E2}{}
\begin{equation}
{\bm \Sigma} 
=
\begin{pmatrix}
\sigma_{11} & \sigma_{12}\\
\sigma_{21} & \sigma_{22}
\end{pmatrix},
\end{equation}
where $\sigma_{12}=\sigma_{21}$. Further, the elements $\sigma_{11}$, $\sigma_{12}$, and $\sigma_{22}$ are expressed as follows (for details, see Appendix \hyperlink{A2}{B}):
\begin{equation*}
\begin{split}
\sigma_{11}&=\left[\frac{1}{2\log 2}\log\left(\frac{p^{*}_{21}}{p^{*}_{21}+1}\right)\right]^{2} \frac{p^{*}_{12}p^{*}_{21}}{p_{12}+p_{21}},\\[3pt]
\sigma_{12}&=\left(\frac{1}{2\log 2}\right)^{2}p^{*}_{12}p^{*}_{21}\log\left(\frac{p^{*}_{21}}{p^{*}_{21}+1}\right)\log\left(\frac{p_{\cdot 1}}{p_{\cdot 1}+1}\right),\\[3pt]
\sigma_{22}&=\left[\frac{1}{2\log 2}\log\left(\frac{p_{\cdot 1}}{p_{\cdot 1}+1}\right)\right]^{2}p_{\cdot 1}p_{\cdot 2}.
\end{split}
\end{equation*}

\hypertarget{S22}{}
\subsection{Predictive monitoring based on the proposed index vector}

Next, we represent the efficacy and toxicity of an experimental treatment, E, and assume that a standard treatment, S, is collected as historical control data. 
The observed outcomes of participants who have entered the trial for E thus far are denoted by ${\bm X}=(X_{11}, X_{12}, X_{21}, X_{22})^{\prime}$, which come from a multinomial distribution with the sum of the current data $n$ and the parameter ${\bm p}_{E}$. In addition, the maximum sample size planned for the entire trial of E is denoted by $N_{max}$.
The multinomial distribution belongs to the exponential family, and its natural conjugate prior density is a Dirichlet distribution. Here, we denote $t=\rm{E,~S}$ as the index treatment. We then introduce a Dirichlet prior for ${\bm p}_{t}=\left(p_{t,11}, p_{t,12}, p_{t,21}, p_{t,22}\right)^{\prime}$,
$$
{\bm p}_{t}|{\bm \alpha}_{t}\sim {\rm Dir}\left({\bm \alpha}_{t}\right),
$$
where ${\bm \alpha}_{t}=(\alpha_{t,11}, \alpha_{t,12}, \alpha_{t,21}, \alpha_{t,22})^{\prime}$ is the vector of hyperparameters such that $\alpha_{t,ij}>0$, $\forall i, j$. From the conjugate analysis, given the observed outcome $\bm x$, we obtain the following posterior distribution of ${\bm p}_{E}$
$$
{\bm p}_{E}|{\bm \alpha}_{E}, {\bm x} \sim {\rm Dir}\left({\bm \alpha}_{E}+{\bm x}\right).
$$
Let ${\bm Y}=(Y_{11}, Y_{12}, Y_{21}, Y_{22})^{\prime}$ be the random efficacy and safety data for potential $m$ future participants, i.e., $n+m=N_{max}$. At the end of the trial, when the data become available for all the $N_{max}$ participants with data ${\bm Y}={\bm y}$ collected, we can construct the following posterior distribution of ${\bm p}_{E}$
$$
{\bm p}_{E}|{\bm \alpha}_{E}, {\bm x}, {\bm y} \sim {\rm Dir}\left({\bm \alpha}_{E}+{\bm x}+{\bm y}\right).
$$
For the interim analysis, data ${\bm y}$ are not observed; thus, we consider a posterior predictive distribution of ${\bm Y}$, given ${\bm \alpha}_{E}$, ${\bm x}$. This is a Dirichlet compound multinomial distribution with the following probability mass function:
$$
f_{m}({\bm y}|{\bm \alpha}_{E}, {\bm x})=\frac{\Gamma(m+1)}{\prod_{i}\prod_{j}\Gamma(y_{ij}+1)}\frac{\Gamma\left(\sum_{i}\sum_{j}\alpha_{E,ij}+x_{ij}\right)}{\Gamma\left(\sum_{i}\sum_{j}\alpha_{E,ij}+x_{ij}+y_{ij}\right)}\prod_{i}\prod_{j}\frac{\Gamma\left(\alpha_{E,ij}+x_{ij}+y_{ij}\right)}{\Gamma\left(\alpha_{E,ij}+x_{ij}\right)},
$$
where $\Gamma (\cdot)$ is a Gamma function. See Appendix \hyperlink{A3}{C} for details on the elicitation of Dirichlet compound multinomial probability mass function. Here, we consider the number of patterns in $\bm Y$. This problem can be formulated to obtain the total number of combinations when the range of each cell is equal to $[0, m]$ under the condition that the sum of the observed frequencies is $m$. In general, the number of solutions is determined by the coefficient of $z^{m}$ in a generating function (\hyperlink{R13}{Hankin, 2007}) $\prod_{i=1}^{T}\sum_{j=0}^{f_{i}}z^{j}$, where 
$$
\sum_{i=1}^{T}y_{i}=m \quad (0\leq y_{i} \leq f_{i}),
$$
and $T$ is the number of cells, i.e., $T=4$, $f_{i}=m$ for all $i$. Notably, the number of patterns increases when there are many $m$ future participants. Based on the data obtained at the end of the trial (i.e., when the maximum sample size is reached), an interim go/no-go decision is made based on the following probability:
\hypertarget{E3}{}
\begin{equation}
{\rm Pr}\left( {\bm \Phi}_{E} - {\bm \Phi}_{S} > {\bm \delta}_{0}|{\bm \alpha}_{S}, {\bm \alpha}_{E},{\bm x}, {\bm y}\right) \geq \lambda,
\end{equation}
where $\lambda$ is a pre-specified probability threshold, ${\bm \delta}_{0}$ is the minimally acceptable increment, and ${\bm \Phi}_{t}$ is obtained by replacing ${\bm p}$ with ${\bm p}_{t}$ from Equation (\hyperlink{E1}{1}). Unless otherwise specified, ${\bm \delta}_{0}$ is considered to be ${\bm 0}=(0, 0)^\prime$. 
To calculate Equation (\hyperlink{E3}{3}), we obtain the asymptotic distribution of the difference between ${\bm \Phi}_{E}$ and ${\bm \Phi}_{S}$. Owing to the properties of the multivariate normal distribution, $\hat{\bm{\Phi}}_{E} - {\hat{\bm \Phi}}_{S} - ({\bm \Phi}_{E} - {\bm \Phi}_{S})$ also has an asymptotically bivariate normal distribution with a zero mean vector and covariate matrix:
\hypertarget{E4}{}
\begin{equation}
\frac{1}{N_{max}+\sum_{i}\sum_{j}\alpha_{E,ij}}{\bm \Sigma}_{E}+\frac{1}{\sum_{i}\sum_{j}\alpha_{S,ij}}{{\bm \Sigma}_S},
\end{equation}
where ${\bm \Sigma}_{t}$ is obtained by replacing ${\bm p}$ with ${\bm p}_{t}$ from Equation (\hyperlink{E2}{2}). To estimate the index vector ${\bm \Phi}_{E}$ and the covariance matrix ${\bm \Sigma}_{E}$, the following Bayesian estimator is plugged in as a consistent estimator:
$$
\hat{p}_{E,ij}=\frac{\alpha_{E,ij}+x_{ij}+y_{ij}}{\sum_{i}\sum_{j}\alpha_{E,ij}+N_{max}}.
$$
However, for ${\bm \Phi}_{S}$ and ${\bm \Sigma}_{S}$, we plugged in $\hat{p}_{S,ij}=\alpha_{S,ij}/\sum_{i}\sum_{j}\alpha_{S,ij}$.
This is because, although the pattern including zero counts is considered in the predictive distribution, the existence of zero counts makes for the estimation of the asymptotic covariance matrix theoretically impossible. Although the range of the proposed indexes is defined as $[0,1]$, this specifies the range of the estimate to be $(0, 1)$.

Subsequently, $I(\cdot)$ is considered an indicator that maps the elements of the subset to one, and all other elements to zero. At the interim monitoring, we can compute the predictive probability that the experimental treatment will be declared sufficiently promising at the scheduled end of the trial using the following equation:
\hypertarget{E5}{}
\begin{equation}
PP=\sum_{\bm y}f_{m}({\bm y}|{\bm \alpha}_{E}, {\bm x})I\left( {\rm Pr}\left( {\bm \Phi}_{E} - {\bm \Phi}_{S} > {\bm \delta}_{0}|{\bm \alpha}_{S}, {\bm \alpha}_{E}, {\bm x}, {\bm y}\right) \geq \lambda \right).
\end{equation}
This is interpreted as the weighted average of the indicators of a successful trial upon the end of the trial, based on the probability of the Dirichlet-multinomial distribution. A high $PP$ indicates that the experimental treatment is likely to be efficacious by the end of the study, given the current data, whereas a low $PP$ suggests that the experimental treatment may not offer sufficient activity. In line with \hyperlink{R16}{Lee and Liu (2008)} and \hyperlink{R21}{Sumbucini (2019)}, the decision rules can be constructed as follows:
\begin{itemize}
\setlength{\parskip}{0cm} 
\setlength{\itemsep}{0cm}
\item if $PP < \theta_{L}$, then stop the trial and declare that the experimental treatment is not sufficiently successful;
\item if $PP > \theta_{U}$, then stop the trial and declare that the experimental treatment is sufficiently successful;
\item otherwise, continue enrolling participants until the maximum sample size is reached.
\end{itemize}
Typically, we choose $\theta_{L}$ as a small positive constant and $\theta_{U}$ as a large positive constant, and their values lie between $0$ and $1$. $PP < \theta_{L}$ indicates that the trial is unlikely to show the desired performance in terms of efficacy and safety at the end of the trial given the current information. However, when $PP > \theta_{U}$, the current data suggest a high probability of concluding that the experimental treatment is efficacious upon the end of the study. Given the nature of single-arm phase II studies, we set $\theta_{U} = 1$ throughout this paper. This is because there is no ethical concern regarding accruing additional participants for a specific situation in which early results appear positive, thereby increasing the precision of the profile of efficacy and safety data (\hyperlink{R11}{Green et al., 2012}, Section 5.1.1).

\hypertarget{S3}{}
\section{Numerical application}

Herein, we demonstrate the process of calculating the predictive probability using both the Monte Carlo and asymptomatic approaches, assuming that the maximum sample size of the study $N_{max}$ is $30$ and that the observed data in Table \hyperlink{T1}{1}(a) are the current data. For the hyperparameter vector of E, Jeffrey's prior is adopted throughout this paper, as it is often used as a non-informative prior and allows for the likelihood to dominate the posterior distribution. Importantly, \hyperlink{R4}{Brutti et al. (2011)} and \hyperlink{R21}{Sumbucini (2019)} derived the reference prior. The Jeffery's prior in the Dirichlet distribution is corresponding to a 1-group reference prior (\hyperlink{R2}{Berger and Bernardo, 1992}). In contrast, we require an informative prior of S, based on a combination of historical control data and clinical experience. Therefore, the values of ${\bm \alpha_{E}}=(0.5, 0.5, 0.5, 0.5)$ and ${\bm \alpha_{S}}=(10, 9, 11, 30)$ are set in this section. For the Monte Carlo approach, we set the number of repeat simulations to 10,000. Hereafter, the distribution using the Monte Carlo approach is referred to as the simulation distribution. 
Here, we assume that ${\bm y}=(0, 3, 2, 0)$ as a result of the future participants. Then, the bivariate index vector of the difference between ${\bm \Phi}_{E}$ and ${\bm \Phi}_{S}$ is estimated $(0.397, 0.106)^\prime$. Figures \hyperlink{F1}{1}(a) and (b) show the simulation distributions of ${\bm \Phi}_{E}$, ${\bm \Phi}_{S}$ and that of the difference using the Monte Carlo approach, whereas Figures \hyperlink{F2}{2}(a) and (b) show those of the asymptotic distribution. 

\begin{center}
\lbrack Figures 1(a) and (b) about here.\rbrack
\end{center}

\begin{center}
\lbrack Figures 2(a) and (b) about here.\rbrack
\end{center}

\noindent
For the asymptotic approach, the estimated covariance matrix of the difference ${\hat{\bm \Sigma}}_{d}$ given by Equation (\hyperlink{E4}{4}), replacing ${\bm p}_t$ with $\hat{\bm p}_{t}$, is as follows:
$$
{\hat{\bm \Sigma}}_{d} 
=
\begin{pmatrix}
0.024&0.010\\
0.010&0.011
\end{pmatrix}.
$$
To shorten the notation, we denote the left side of Equation (\hyperlink{E3}{3}) by $B(\bm Y)$ in this section. Then, $B(\bm Y)$ is calculated to be $0.8457$ and $0.8378$ using the Monte Carlo and asymptotic approaches, respectively.

Next, we consider all the possible responses $\bm Y$, including ${\bm y}=(0, 3, 2, 0)$, among the potential $m=5$ future participants; that is, all the possible entries of $2 \times 2$ square contingency table consistent with a total equal to $5$, i.e., $N_{max}=30$. When the threshold $\lambda$ is set to $0.80$, Tables \hyperlink{T3}{3} and \hyperlink{T4}{4} present the proposed predictive probability based on Monte Carlo and asymptotic approaches, respectively. 

\begin{center}
\lbrack Tables 3 and 4 about here.\rbrack
\end{center}

\noindent
In Tables \hyperlink{T3}{3} and \hyperlink{T4}{4}, we enumerate all the possible values of $\bm Y$, and the second and sixth columns provide its posterior predictive probability computed by $f_{m}(\bm y)$. For each item, we show its posterior predictive probability of declaring the treatment promising at the end of the trial for both the Monte Carlo and asymptotic approaches. A further column provides the value of the indicator, which is equal to $1$ if the corresponding posterior probability $B(\bm Y)$ exceeds $\lambda$, otherwise $0$. By simply summarizing it, we obtain the predictive probability of interest. Thus, the values of $PP$ given by Equation (\hyperlink{E5}{5}) are 0.907 for both the Monte Carlo and asymptotic approaches. For the data in Table \hyperlink{T1}{1}(b), we can obtain the values of $PP$ as 0.743 and 0.732 for the Monte Carlo and asymptotic approaches, respectively.

\hypertarget{S4}{}
\section{Properties of the asymptotic approach}

This section presents an approximation of the asymptotic approach. To evaluate the performance of the asymptotic approach used in this paper, various probability scenarios are considered. First, the null  ($H_{0}$) and alternative ($H_{1}$) hypotheses are set to visualize the behavior of each approach as presented by these specific hypotheses:
$$
H_{0}:~{\bm \Phi}_{E}={\bm \Phi}_{S}\quad {\rm vs}\quad H_{1}:~{\bm \Phi}_{E}\not={\bm \Phi}_{S}.
$$
We assume that $\bm{P}_{E}=(0.15, 0.30, 0.15, 0.40)$ under $H_{0}$, $\bm{P}_{E}=(0.10, 0.50, 0.05, 0.35)$ under $H_{1}$, and the threshold $\lambda$ is set to $0.80$.
Figure \hyperlink{F3}{3} illustrates the predictive probabilities with various values of $n$ for the number of current data of E and ${\bm \alpha}_{S}$. In total, nine patterns of ${\bm \alpha}_{S}$ are obtained by setting $\sum_{i}\sum_{j}\alpha_{S,ij}$ from $60$ to $220$ by $20$, with ${\bm p}_{S}=(0.15, 0.30, 0.15, 0.40)$. In addition, four patterns are obtained in terms of the current amount of data $n$ from $20$ to $80$ by $20$, and the remaining data of E ($m$) is set to $20$ participants. Figures \hyperlink{F3}{3}(a) and \hyperlink{F3}{3}(b) show the results under $H_{0}$ and $H_{1}$, respectively. 

\begin{center}
\lbrack Figures 3(a) and (b) about here.\rbrack
\end{center}

\noindent
As seen in Figures \hyperlink{F3}{3}(a) and (b), even if the number of current participants $n$ in the clinical trial is $20$, the asymptotic approach can well approximate the results of the Monte Carlo approach. However, because the approximation depends on the precision, i.e., variance of the distribution, care should be taken when the amount of information contributing to S, such as the sample size of historical control data, is relatively small. This is because the probability of success may not increase above a certain level irrespective of  the size of the single-arm phase II study. This is evident from the asymptotic distribution of the difference given in Equation (\hyperlink{E4}{4}). Therefore, optimization of the tuning parameter is important (see Section \hyperlink{S6}{6}).

\hypertarget{S5}{}
\section{Operating characteristics of the proposed design}
In this section, we use a simulation to empirically investigate the performance of the proposed Bayesian interim procedure based on an asymptotic approach in terms of frequentist operating characteristics.
As discussed in Section \hyperlink{S2}{2}, we set $\theta_{U}=1$ such that the trial can be terminated only because of low efficacy or excessive toxicity and also set $\theta_{L}=0.001$. In addition, we set ${\bm \alpha}_{E}=(0.5, 0.5, 0.5, 0.5)$ as Jeffrey's prior to have a vague prior. For the hyperparameter of S, we set ${\bm \alpha}_{S}=(30, 60, 30, 80)$ with $\sum_{i}\sum_{j}\alpha_{S,ij}=200$ as an informative prior based on the results shown in Section \hyperlink{S4}{4}.  For reference, we also describe that the widths of the 95\% credible intervals for the marginal distributions of $\Phi_{eff}$ and $\Phi_{tox}$ are estimated to be $0.195$ and $0.134$, respectively, according to the asymptotic approach. Furthermore, \hyperlink{R30}{Thall and Simon (1994b)}, \hyperlink{R38}{Zohar et al. (2008)}, and \hyperlink{R14}{Hirakawa et al. (2018)} offer examples on the width of the credible interval.
The threshold $\lambda$ is set to $0.80$. We assume that the first interim analysis is conducted after observing $N_{min}=10$ participants, and subsequently, data are monitored using cohorts of size one or five until the maximum sample size $N_{max}=40$ is reached. To evaluate the performance of the designs, we considered the following three metrics: 
(i) The percentage of early terminations (PET) is defined as the percentage of trials that are terminated early.
(ii) The percentage of rejection of the null hypothesis (PRN) is defined as the percentage of the simulated trials in which $H_{0}$ is rejected, i.e., the frequency of simulated trials that are not stopped before the maximum sample size is reached. The PRN is the type I error rate (or power) when $H_{0}$ is (or is not) true, or the percentage of claims that the experimental treatment is effective.
(iii) The actual sample size is defined as the average sample size (ASS) actually used in 10,000 simulated trials.

\subsection{Simulation study 1}
We first evaluate the operating characteristics of the proposed Bayesian interim procedure for comparison with \hyperlink{R21}{Sambucini (2019)} and to evaluate a realistic and practical setting with reference to the settings in \hyperlink{R21}{Sambucini (2019)} and \hyperlink{R37}{Zhou et al. (2017)}. Because the proposed method is dependent on the precision of S, it can be interpreted as a comparison with the case in which the design of \hyperlink{R21}{Sambucini (2019)} is applied when the relevant information, such as ${\bm \alpha}_{S}$, is available before the onset of the trial. Table \hyperlink{T5}{5} lists the operating characteristics of the proposed designs, summarized by the empirical probabilities of PET, PRN, and ASS. 

\begin{center}
\lbrack Table 5 about here.\rbrack
\end{center}

\noindent
Consider the scenarios 1 and 2 presented in Table \hyperlink{T5}{5}. Both are characterized by the true joint probability of efficacy and safety $p_{12}$ being equal; however, the true joint probability of non-efficacy and toxicity $p_{21}$ and the marginal probability of toxicity $p_{\cdot 1}$ are different. Thus, compared to scenario 2, scenario 1 is a slightly better setting.  As a result, scenario 1 has a lower PET and higher PRN and ASS than scenario 2 for each cohort size. 
In addition, scenarios 1 and 4 should have almost the same setting, as the true ratios of $p_{12}$ and $p_{21}$ are the same and true $p_{\cdot 1}$ is also the same. Consequently, both groups are characterized as having almost the same PET, PRN, and ASS for each cohort size. In contrast, scenario 5 is clearly inferior to scenario 1, because both have the same true $p_{21}$ and true $p_{\cdot 1}$, but the true $p_{12}$ is different. Thus, scenario 1 has a far lower PET and higher PRN and ASS than scenario 5 for each cohort size. 
Furthermore, consider the last two scenarios presented in Table 5, i.e., scenarios 9 and 10. Scenario 10 is a slightly more promising situation with same true $p_{\cdot 1}$ and different true ratio of $p_{12}$ and $p_{21}$, and so do scenarios 6 and 7. Consequently, scenario 9 has a higher PET and lower PRN and ASS than scenario 10. Similarly, scenario 6 has a higher PET and lower PRN and ASS than scenario 7. 
For the cohort size, ASS increases and PET decreases overall in cohort size 5, compared with that in cohort size 1.
In our simulation study, comparisons with \hyperlink{R21}{Sambucini (2019)} yield similar results.

\subsection{Simulation study 2}
For the motivations presented in Section \hyperlink{S1}{1}, we evaluate the behavior of the proposed method in scenarios where, by definition, \hyperlink{R21}{Sambucini (2019)} exhibits the same operating characteristics. 

\begin{center}
\lbrack Table 6 about here.\rbrack
\end{center}

\noindent
Table \hyperlink{T6}{6} lists the operating characteristics of the proposed designs, summarized by the empirical probabilities of PET, PRN, and ASS.
As expected, PET tends to decrease from scenarios 1 to 4, and PRN and ASS tend to increase. In other words, we are able to capture the structure of cases with toxicity without efficacy. For cohort size, as in simulation study 1, ASS increases and PET decreases overall in cohort size 5, compared with that in cohort size 1.

\hypertarget{S6}{}
\section{Discussion}
We proposed a new design for a Bayesian single phase II study with both efficacy and safety endpoints. Specifically, an index vector was employed that appropriately summarized the situation by considering the structure in the case where toxicity occurred but efficacy was not achieved, accounted for the uncertainty in the historical control data through the use of prior distribution.
Furthermore, the operating characteristics of the proposed method were demonstrated by a comprehensive simulation study. Notably, the proposed design yielded desirable operating characteristics under various settings. Consequently, the proposed method makes appropriate interim go/no-go decision. 

In Section \hyperlink{S5}{5}, the tuning parameters were set to the same values as in \hyperlink{R21}{Sambucini (2019)} for comparison; however, the optimization of these tuning parameters will be important for practical use. Therefore, we also illustrate the calibration of the design thresholds to ensure desirable frequentist operating characteristics under the selected scenario of interest. Similar to \hyperlink{R36}{Yin et al. (2012)}, considering the null cases, all entries with type I error rates are 10\% or less in our study, for which the paired values of $(\lambda, \theta_{L})$ satisfy our requirements. Moreover, under the alternative cases, we need to find the paired value that provides a maximum power of more than 80\%.
Given that we follow the aforementioned setting of parameters with a cohort size of five participants other than the pair of $(\lambda, \theta_{L})$, we conduct 10,000 simulated clinical trials. For each of the paired values of  $(\lambda, \theta_{L})$, we obtain the type I error and power listed in Table \hyperlink{T7}{7}. 

\begin{center}
\lbrack Table 7 about here.\rbrack
\end{center}

\noindent
In this case, the type I error rate is guaranteed to be less than $0.10$ except for the combinations $(\lambda, \theta_{L})=(0.70, 0.001)$ and $(\lambda, \theta_{L})=(0.70, 0.01)$. One of these combinations can be selected to maximize the power using the boundary line of the staircase curve. In this case, it is appropriate to set $(\lambda, \theta_{L})=(0.75, 0.01)$, which is  highlighted in gray at the time of designing a clinical trial.

Although the proposed method is considered to be a powerful method for making appropriate decisions by optimizing the tuning parameters, a few limitations exist. First, as pointed out by \hyperlink{R24}{Saville, et al. (2014)}, a problem with computational burden still remains. For example, as discussed in Section \hyperlink{S2}{2}, the number of patterns increased because of the considerations of efficacy and safety, and predictive probability calculation is exhaustive. To circumvent this issue, in addition to the Monte Carlo approach, we proposed an asymptotic approach and evaluated its performance. Although the performance of the asymptotic approach was good, these results were limited to our simulation pattern, and thus, we should be careful about generalizing it. However, in practical use, the accuracy of the approximation to the Monte Carlo approach may not have a major impact in a well-designed clinical trial that is based on an asymptotic approach with sound operating characteristics.
Second, as explained in Section \hyperlink{S4}{4}, the probability of success may not increase above a certain level regardless of the size of the single-arm phase II study. Depending on the accuracy, the proposed predictive probability may be saturated by the location parameter. Thus, optimizing the tuning parameter is essential, and the aforementioned calibration is strongly recommended. Notably, in the Bayesian framework, the FDA (\hyperlink{R33}{Guidance industry, 2020}) also states that it is possible to use simulations to estimate the frequentist operating characteristics of power and type I error probability. It also states that decision criteria can be chosen to provide type I error control at a specified level.
Additionally, the proposed method requires the use of joint distribution of efficacy and toxicity. It may be more difficult to obtain historical control data compared to cases where the results are referred to as marginal distributions, i.e., aggregated data, for efficacy and toxicity. Therefore, taking into account the issue of a small sample size, it is important to make an effort to mature the data when the historical control data are generally small e.g., by accessing the individual patient data. Further, it is desirable to elaborate on the data, e.g., by considering access to real-world data sources. Lastly, regarding the stopping boundary, although it is important that the go/no-go decision boundary can be enumerated before the onset of the trial, the proposed method cannot calculate it due to its dependency on the joint structure of the interim data. This issue was also highlighted by \hyperlink{R22}{Sambucini (2021a)}. Thus, statisticians must calculate the predictive probability for each interim monitoring time point.

Regarding the interpretation of the bivariate index, as this is a new index, and no clinical results are yet available based on it, we cannot rule out the possibility that the index itself is difficult to interpret. However, in the context of the predictive probability, \hyperlink{R24}{Saville, et al. (2014)} states that clinical trial designs using predictive probability for interim monitoring do not claim efficacy using predictive probability. Rather, the claim of treatment benefit is based on either Bayesian posterior probability or frequentist criteria. Similarly, the proposed method is used only as a decision-making framework and is not intended to evaluate the magnitude of the treatment effect. 
In this respect, it is important to comprehensively evaluate the study results by considering, e.g., the posterior distribution of observed response rate and the frequency of adverse events of special interest.

As discussed in Section \hyperlink{S1}{1}, for situations where efficacy and safety are co-primary, it is natural to assume that the risk--benefit do not depend on a diagonal cell. However, 
in situations where the agent is valuable, despite the joint probability of a responder who simultaneously experiences adverse toxicity being high, the proposed method cannot capture its characteristics. Practically, it may be desirable to apply the proposed method in situations where no toxicity should be assumed. For instance, the presence of toxicity of interest is set as a quite critical event, e.g., adverse events such as dose-limiting toxicity or leading to death.

In this paper, we evaluated the operating characteristics of cohorts of one and five participants. Of course, although the performance was generally better with a cohort size of one participant, it was confirmed that the performance was not so inferior to that of five participants. However, provided that the clinical trial is statistically well designed, it is also important to preserve the study integrity in conducting clinical trials. If the cohort size is set at one participant, there is a concern that it may lead to bias in the efficacy and safety evaluation compared to the case with five participants. For example, it is impossible to deny any bias that may occur when the person in charge of the study knows that the study will be stopped if only one response is obtained or if one toxicity is observed.
Thus, it may be appropriate to set the cohort size to e.g., within five participants when conducting the proposed method, provided that the performance is not so inferior.

Finally, in the proposed method, the predictive probability is evaluated based on the joint distribution of the efficacy and toxicity indexes. However, an evaluation with emphasis on either efficacy or toxicity should be considered instead. In this case, a weight is given to the indexes of each efficacy and toxicity, and a one-dimensional index can be constructed as the weighted values. Furthermore, the proposed method assumes that the endpoints can be observed quickly enough, such that adaptive go/no-go decisions can be made in a timely fashion at each interim monitoring. Thus, we could postpone the accrual at the interim to wait for the data to be cleared. However, this approach is typically infeasible in practice as repeatedly suspending participant accrual is impractical and may lead to unacceptably long trials. For example, if confirmation is required to determine the best overall response according to RECIST for efficacy evaluation, this can lead delayed response. To address these issues, in the context of phase I study, \hyperlink{R39}{Asakawa and Hamada (2013)} proposed an approach using unconfirmed early responses as the surrogate efficacy outcome for the confirmed outcome. In addition, \hyperlink{R7}{Cai et al. (2014)} proposed a Bayesian trial design to allow continuous monitoring of phase II clinical trials in the presence of delayed responses. Therefore, it is desirable to exploit a new design to allow for the delayed responses in the proposed design.

\

\

\noindent
{\large\textit{Acknowledgements}}\\
The authors thank Takashi Asakawa, Kosei Tajima, and Yuki Nakagawa at Chugai Pharmaceutical Co., Ltd. for their helpful comments and suggestions.

\

\noindent
{\large\textit{Author Contributions}}\\
Takuya Yoshimoto contributed to the study conception and design. The analysis was performed by Takuya Yoshimoto, and all authors checked the results. The first draft of the manuscript was written by Takuya Yoshimoto, and all authors commented on previous versions of the manuscript. All the authors have read and approved the final version of the manuscript.

\

\noindent
{\large\textit{Funding}}\\
This research received no specific grant from funding agencies in the public, commercial, or non-for-profit sectors.

\

\noindent
{\large\textit{Conflict of Interest}}\\
Takuya Yoshimoto is an employee of Chugai Pharmaceutical Co., Ltd. The other authors have no conflicts of interest to declare.

\

\noindent
{\large\textit{Data Availability Statement}}\\
All source code can be requested from the corresponding author.

\

\noindent
{\large\textit{ORCID}}\\
Takuya Yoshimoto, \url{https://orcid.org/0000-0002-8747-1395}

\newpage
\hypertarget{A1}{}
\subsection*{Appendix A: Elicitation of the proposed indexes}
We denote conditional probabilities ${\bm p}^{c} = \left(p_{12}^{*}, p_{21}^{*}\right)^{\prime}$, ${\bm q}^{c} = \left(q_{12}^{*}, q_{21}^{*}\right)^{\prime}$, and marginal probabilities ${\bm p}^{m} = \left(p_{\cdot 1}, p_{\cdot 2}\right)^{\prime}$, ${\bm q}^{m} = \left(q_{\cdot 1}, q_{\cdot 2}\right)^{\prime}$.
For the proposed approach, we can consider the following ${\rm JSD}({\bm p}^{c}:{\bm q}^{c})$ and ${\rm JSD}({\bm p}^{m}:{\bm q}^{m})$, respectively:
\begin{equation*}
\begin{split}
{\rm JSD}({\bm p}^{c}:{\bm q}^{c})&=\displaystyle\frac{1}{2}\mathop{\sum\sum}_{i \neq j}\left(p^{*}_{ij}\log\frac{2p^{*}_{ij}}{p^{*}_{ij}+q^{*}_{ij}}+q^{*}_{ij}\log\frac{2q^{*}_{ij}}{p^{*}_{ij}+q^{*}_{ij}}\right),\\[3pt]
{\rm JSD}({\bm p}^{m}:{\bm q}^{m})&=\displaystyle\frac{1}{2}\sum_{i}\left(p_{\cdot i}\log\frac{2p_{\cdot i}}{p_{\cdot i}+q_{\cdot i}}+q_{\cdot i}\log\frac{2q_{\cdot i}}{p_{\cdot i}+q_{\cdot i}}\right).
\end{split}
\end{equation*}
For the probability distribution ${\bm q}$ of the worst case, $q^{*}_{12}=0$  (then $q^{*}_{21}=1$) and $q_{\cdot 1}=1$ (then $q_{\cdot 2}=0$), ${\rm JSD}({\bm p}^{c}:{\bm q}^{c})$ and ${\rm JSD}({\bm p}^{m}:{\bm q}^{m})$ are expressed as follows:
\begin{equation*}
\begin{split}
{\rm JSD}({\bm p}^{c}:{\bm q}^{c})&=\displaystyle\frac{1}{2}\left(p^{*}_{12}\log 2 + p^{*}_{21}\log\frac{2p^{*}_{21}}{p^{*}_{21}+1}+\log\frac{2}{p^{*}_{21}+1}\right),\\[3pt]
{\rm JSD}({\bm p}^{m}:{\bm q}^{m})&=\displaystyle\frac{1}{2}\left(p_{\cdot 2}\log 2 + p_{\cdot1}\log\frac{2p_{\cdot1}}{p_{\cdot1}+1}+\log\frac{2}{p_{\cdot1}+1}\right).
\end{split}
\end{equation*}
Then, divide by $\log 2$ to standardize the maximum value to $1$. Consequently, we can obtain the following indexes:
\begin{equation*}
\begin{split}
\displaystyle \Phi_{eff}&=\displaystyle\frac{1}{2\log 2}\left(p^{*}_{12}\log 2 + p^{*}_{21}\log\frac{2p^{*}_{21}}{p^{*}_{21}+1}+\log\frac{2}{p^{*}_{21}+1}\right),\\[3pt]
\displaystyle \Phi_{tox}&=\displaystyle\frac{1}{2\log 2}\left(p_{\cdot 2}\log 2 + p_{\cdot1}\log\frac{2p_{\cdot1}}{p_{\cdot1}+1}+\log\frac{2}{p_{\cdot1}+1}\right).
\end{split}
\end{equation*}

\

\hypertarget{A2}{}
\subsection*{Appendix B: Asymptotic variance--covariance matrix of the bivariate index vector}
Using multivariate central limit theorem and Slutsky's theorem, with the sample size of $N$, $\sqrt{N}\left(\hat{\bm{p}}-\bm{p}\right)$ has asymptotically a multivariate normal distribution with zero mean and covariance matrix $\bm{{\rm Diag}}(\bm p) - {\bm p}{\bm p^{\prime}}$, where $\hat{\bm{p}}$ is consistent estimator of $\bm{p}$ and $\bm{{\rm Diag}}(\bm p)$ is diagonal matrix with the elements of $\bm{p}$ on the main diagonal (see, e.g., \hyperlink{R1}{Agresti, 2013}, p.590). Thus, $\hat{\bm{\Phi}}$ is given by $\bm {\Phi}$, where $\Phi_{eff}$ and $\Phi_{tox}$ are replaced by $\hat{\Phi}_{eff}$ and $\hat{\Phi}_{tox}$. Let $\left({\partial {\bm \Phi}}/{\partial {\bm p}^\prime} \right)$ denote $2 \times 4$ matrix as follows:
\begin{equation*}
\begin{split}
\left(\frac{\partial {\bm \Phi}}{\partial {\bm p}^\prime} \right)
&=
\begin{pmatrix}
\displaystyle{\partial \Phi_{eff}}/{\partial p_{11}} & \displaystyle{\partial \Phi_{eff}}/{\partial p_{12}} & \displaystyle{\partial \Phi_{eff}}/{\partial p_{21}} & \displaystyle{\partial \Phi_{eff}}/{\partial p_{22}} \\
\displaystyle{\partial \Phi_{tox}}/{\partial p_{11}} & \displaystyle{\partial \Phi_{tox}}/{\partial p_{12}} & \displaystyle{\partial \Phi_{tox}}/{\partial p_{21}} & \displaystyle{\partial \Phi_{tox}}/{\partial p_{22}}
\end{pmatrix}\\[3pt]
&=
{\renewcommand\arraystretch{2.5}
\begin{pmatrix}
0 & \displaystyle\frac{-p_{21}}{2(p_{12}+p_{21})^2\log 2}\log\frac{p^{*}_{21}}{p^{*}_{21}+1} & \displaystyle\frac{p_{12}}{2(p_{12}+p_{21})^2\log 2}\log\frac{p^{*}_{21}}{p^{*}_{21}+1} & 0 \\
\displaystyle\frac{1}{2\log 2}\log\frac{p_{\cdot1}}{p_{\cdot1}+1} & 0 & \displaystyle\frac{1}{2\log 2}\log\frac{p_{\cdot1}}{p_{\cdot1}+1} & 0
\end{pmatrix}
.}
\end{split}
\end{equation*}
When $N$ approaches infinity, the estimated index vector can be approximated as
$$
\hat{\bm{\Phi}} = {\bm \Phi} + \left(\frac{\partial {\bm \Phi}}{\partial {\bm p}^\prime} \right)
\left(\hat{\bm{p}} - \bm{p}\right)+o\left(\|\hat{\bm{p}} - \bm{p}\|\right),
$$
where $o\left(\|\hat{\bm{p}} - \bm{p}\|\right)$ tends toward $\left(0, 0\right)^{\prime}$. Using the delta method (see \hyperlink{R1}{Agresti, 2013}, Section 16.1), $\sqrt{N}\left(\hat{\bm{\Phi}}-\bm{\Phi}\right)$ has an asymptotically bivariate normal distribution with zero mean and covariance matrix:
$$
{\bm \Sigma} = \left(\frac{\partial {\bm \Phi}}{\partial {\bm p}^\prime} \right)
\left(\bm{{\rm Diag}}(\bm p) - {\bm p}{\bm p^{\prime}}\right)
\left(\frac{\partial {\bm \Phi}}{\partial {\bm p}^\prime} \right)^{\prime}
=
\begin{pmatrix}
\sigma_{11} & \sigma_{12}\\
\sigma_{21} & \sigma_{22}
\end{pmatrix},
$$
where $\sigma_{12}=\sigma_{21}$. The elements $\sigma_{11}$, $\sigma_{12}$ and $\sigma_{22}$ are calculated as follows:
\begin{equation*}
\begin{split}
\sigma_{11}&=\left[\frac{1}{2\log 2}\log\left(\frac{p^{*}_{21}}{p^{*}_{21}+1}\right)\right]^{2} \frac{p^{*}_{12}p^{*}_{21}}{p_{12}+p_{21}},\\[3pt]
\sigma_{12}&=\left(\frac{1}{2\log 2}\right)^{2}p^{*}_{12}p^{*}_{21}\log\left(\frac{p^{*}_{21}}{p^{*}_{21}+1}\right)\log\left(\frac{p_{\cdot 1}}{p_{\cdot 1}+1}\right),\\[3pt]
\sigma_{22}&=\left[\frac{1}{2\log 2}\log\left(\frac{p_{\cdot 1}}{p_{\cdot 1}+1}\right)\right]^{2}p_{\cdot 1}p_{\cdot 2}.
\end{split}
\end{equation*}
In addition, based on the large-sample properties of Bayes procedures (\hyperlink{R35}{Wasserman, 2004}, Section 11.5), we can regard the above asymptotic distribution as an asymptotic posterior distribution of ${\bm \Phi}$.

\

\hypertarget{A3}{}
\subsection*{Appendix C: Dirichlet compound multinomial probability mass function}
Let $\Omega$ denote the sample space defined by
$$
\left\{{\bm p}_{E};~p_{E,ij} \in [0,1], ~i, j=1,2, ~\sum_{i}\sum_{j}p_{E,ij}=1\right\}.
$$
With reference to \hyperlink{R32}{Thors\'en (2014)}, Dirichlet compound multinomial probability mass function is thoroughly calculated as follows:
\begin{equation*}
\begin{split}
f_{m}({\bm y}|{\bm \alpha}_{E}, {\bm x})&=\int_{\Omega}f({\bm y}|{\bm p}_{E})f({\bm p}_{E}|{\bm \alpha}_{E}, {\bm x})d{\bm p}_{E}\\
&=\int_{\Omega}\frac{m!}{\prod_{i}\prod_{j}y_{ij}!}\prod_{i}\prod_{j}p^{y_{ij}}_{E,ij}\frac{\Gamma \left(\sum_{i}\sum_{j}\alpha_{E,ij}+x_{ij}\right)}{\prod_{i}\prod_{j}\Gamma\left(\alpha_{E,ij}+x_{ij}\right)}\prod_{i}\prod_{j}p^{\alpha_{E,ij}+x_{ij}-1}_{E,ij}d{\bm p}_{E}\\
&=\frac{m!}{\prod_{i}\prod_{j}y_{ij}!}\frac{\Gamma \left(\sum_{i}\sum_{j}\alpha_{E,ij}+x_{ij}\right)}{\prod_{i}\prod_{j}\Gamma\left(\alpha_{E,ij}+x_{ij}\right)}\int_{\Omega}\prod_{i}\prod_{j}p^{\alpha_{E,ij}+x_{ij}+y_{ij}-1}_{E,ij}d{\bm p}_{E}.
\end{split}
\end{equation*}
We then consider a Dirichlet distribution with parameter ${\bm \alpha}_{E}+{\bm x}+{\bm y}$. From the normalization constant, we obtain
\begin{equation*}
\int_{\Omega}\prod_{i}\prod_{j}p^{\alpha_{E,ij}+x_{ij}+y_{ij}-1}_{E,ij}d{\bm p}_{E}=
\frac{\prod_{i}\prod_{j}\Gamma \left(\alpha_{E,ij}+x_{ij}+y_{ij}\right)}{\Gamma\left(\sum_{i}\sum_{j}\alpha_{E,ij}+x_{ij}+y_{ij}\right)}.
\end{equation*}
Accordingly, we can obtain the following Dirichlet compound multinomial probability mass function:
$$
f_{m}({\bm y}|{\bm \alpha}_{E}, {\bm x})=\frac{\Gamma(m+1)}{\prod_{i}\prod_{j}\Gamma(y_{ij}+1)}\frac{\Gamma\left(\sum_{i}\sum_{j}\alpha_{E,ij}+x_{ij}\right)}{\Gamma\left(\sum_{i}\sum_{j}\alpha_{E,ij}+x_{ij}+y_{ij}\right)}\prod_{i}\prod_{j}\frac{\Gamma\left(\alpha_{E,ij}+x_{ij}+y_{ij}\right)}{\Gamma\left(\alpha_{E,ij}+x_{ij}\right)}.
$$

\newpage
\hypertarget{T1}{}
\noindent
{\bf{Table 1:}}
Artificial data on efficacy and toxicity in the interim stage.

\

\renewcommand{\arraystretch}{1.3}
\begin{table}[htbp]
\begin{minipage}[c]{0.5\hsize}
\centering
\begin{tabular} {lccc} 
(a)&&&\\
\hline
& \multicolumn{2}{c}{Toxicity}  &\\ \cline{2-3}
Efficacy~~ & ~~Yes~~ & ~~No~~ &~~Total~~ \\\hline
Yes       & 5     & 10 & 15\\
No        & 0     & 10 & 10\\\hline
Total     & 5     & 20 & 25\\\hline
\end{tabular}
\end{minipage}
\begin{minipage}[c]{0.5\hsize}
\centering
\begin{tabular} {lccc} 
(b)&&&\\
\hline
& \multicolumn{2}{c}{Toxicity}  &\\ \cline{2-3}
Efficacy~~ & ~~Yes~~ & ~~No~~ &~~Total~~ \\\hline
Yes       & 0     & 10 & 10\\
No        & 5     & 10 & 15\\\hline
Total     & 5     & 20 & 25\\\hline
\end{tabular}
\end{minipage}
\end{table}

\

\

\hypertarget{T2}{}
\noindent
{\bf{Table 2:}}
Square contingency table with binary endpoints representing efficacy and toxicity.

\

\renewcommand{\arraystretch}{1.3}
\begin{table}[htbp]
\centering
\begin{tabular} {lccc} 
\hline
& \multicolumn{2}{c}{Toxicity}   &\\ \cline{2-3}
Efficacy~~ & ~~Yes~~     & ~~No~ ~ &~~Total~~ \\\hline
Yes       & $p_{11}$        & $p_{12}$ & $p_{1\cdot}$\\
No        & $p_{21}$         & $p_{22}$ & $p_{2\cdot}$ \\\hline
Total     & $p_{\cdot 1}$  & $p_{\cdot 2}$ & $1$ \\\hline
\end{tabular}
\end{table}

\newpage
\hypertarget{T3}{}
\noindent
{\bf{Table 3:}}
Detailed calculations to obtain the predictive probability using the Monte Carlo approach with 10,000 times repeated simulations when ${\bm \alpha}_{E}=(0.5, 0.5, 0.5, 0.5)$, ${\bm \alpha}_{S}=(10, 9, 11, 30)$, ${\bm x}=(5, 10, 0, 10)$, $\lambda=0.80$, $N_{max}=30$.

\

\renewcommand{\arraystretch}{1.3}
\begin{table}[htbp]
\centering
\begin{tabular} {lccclccc} 
\hline
$\bm y$ & $f_{m}({\bm y}|{\bm \alpha}_{E}, {\bm x})$ & $B(\bm Y)$ & $I(B(\bm Y)>\lambda)$ &  $\bm y$ & $f_{m}({\bm y}|{\bm \alpha}_{E}, {\bm x})$ & $B(\bm Y)$ & $I(B(\bm Y)>\lambda)$ \\\hline
$(5,0,0,0)$ & $0.0010619$ & $0.5284$ & $0$ &$(1,2,1,1)$& $0.0102602$ & $0.8533$ & $1$\\
$(4,1,0,0)$ & $0.0058683$ & $0.6469$ & $0$ &$(0,3,1,1)$& $0.0077729$ & $0.9129$ & $1$\\
$(3,2,0,0)$ & $0.0158789$ & $0.7551$ & $0$ &$(2,0,2,1)$& $0.0004142$ & $0.6377$ & $0$\\
$(2,3,0,0)$ & $0.0264649$ & $0.8508$ & $1$ &$(1,1,2,1)$& $0.0013383$ & $0.7671$ & $0$\\
$(1,4,0,0)$ & $0.0274828$ & $0.9138$ & $1$ &$(0,2,2,1)$& $0.0013991$ & $0.8443$ & $1$\\
$(0,5,0,0)$ & $0.0144909$ & $0.9574$ & $1$ &$(1,0,3,1)$& $0.0001062$ & $0.6508$ & $0$\\
$(4,0,1,0)$ & $0.0002794$ & $0.5366$ & $0$ &$(0,1,3,1)$& $0.0002028$ & $0.7461$ & $0$\\
$(3,1,1,0)$ & $0.0013808$ & $0.6548$ & $0$ &$(0,0,4,1)$& $0.0000169$ & $0.6464$ & $0$\\
$(2,2,1,0)$ & $0.0031758$ & $0.7649$ & $0$ &$(3,0,0,2)$& $0.0158789$ & $0.7551$ & $0$\\
$(1,3,1,0)$ & $0.0040715$ & $0.8435$ & $1$ &$(2,1,0,2)$& $0.0666916$ & $0.8419$ & $1$\\
$(0,4,1,0)$ & $0.0024984$ & $0.9101$ & $1$ &$(1,2,0,2)$& $0.1179928$ & $0.9145$ & $1$\\
$(3,0,2,0)$ & $0.0000986$ & $0.5268$ & $0$ &$(0,3,0,2)$& $0.0893885$ & $0.9562$ & $1$\\
$(2,1,2,0)$ & $0.0004142$ & $0.6387$ & $0$ &$(2,0,1,2)$& $0.0031758$ & $0.7643$ & $0$\\
$(1,2,2,0)$ & $0.0007329$ & $0.7556$ & $0$ &$(1,1,1,2)$& $0.0102602$ & $0.8531$ & $1$\\
$(0,3,2,0)$ & $0.0005552$ & $0.8441$ & $1$ &$(0,2,1,2)$& $0.0107266$ & $0.9165$ & $1$\\
$(2,0,3,0)$ & $0.0000329$ & $0.5395$ & $0$ &$(1,0,2,2)$& $0.0007329$ & $0.7546$ & $0$\\
$(1,1,3,0)$ & $0.0001062$ & $0.6528$ & $0$ &$(0,1,2,2)$& $0.0013991$ & $0.8437$ & $1$\\
$(0,2,3,0)$ & $0.0001110$ & $0.7598$ & $0$ &$(0,0,3,2)$& $0.0001110$ & $0.7558$ & $0$\\
$(1,0,4,0)$ & $0.0000089$ & $0.5334$ & $0$ &$(2,0,0,3)$& $0.0264649$ & $0.8507$ & $1$\\
$(0,1,4,0)$ & $0.0000169$ & $0.6495$ & $0$ &$(1,1,0,3)$& $0.0855020$ & $0.9148$ & $1$\\
$(0,0,5,0)$ & $0.0000014$ & $0.5246$ & $0$ &$(0,2,0,3)$& $0.0893885$ & $0.9562$ & $1$\\
$(4,0,0,1)$ & $0.0058683$ & $0.6469$ & $0$ &$(1,0,1,3)$& $0.0040715$ & $0.8428$ & $1$\\
$(3,1,0,1)$ & $0.0289963$ & $0.7623$ & $0$ &$(0,1,1,3)$& $0.0077729$ & $0.9123$ & $1$\\
$(2,2,0,1)$ & $0.0666916$ & $0.8419$ & $1$ &$(0,0,2,3)$& $0.0005552$ & $0.8404$ & $1$\\
$(1,3,0,1)$ & $0.0855020$ & $0.9149$ & $1$ &$(1,0,0,4)$& $0.0274828$ & $0.9137$ & $1$\\
$(0,4,0,1)$ & $0.0524672$ & $0.9575$ & $1$ &$(0,1,0,4)$& $0.0524672$ & $0.9574$ & $1$\\
$(3,0,1,1)$ & $0.0013808$ & $0.6548$ & $0$ &$(0,0,1,4)$& $0.0024984$ & $0.9088$ & $1$\\
$(2,1,1,1)$ & $0.0057993$ & $0.7600$ & $0$ &$(0,0,0,5)$& $0.0144909$ & $0.9573$ & $1$\\\hline

\end{tabular}
\end{table}

\newpage
\hypertarget{T4}{}
\noindent
{\bf{Table 4:}}
Detailed calculations to obtain the predictive probability using the asymptotic approach when ${\bm \alpha}_{E}=(0.5, 0.5, 0.5, 0.5)$, ${\bm \alpha}_{S}=(10, 9, 11, 30)$, ${\bm x}=(5, 10, 0, 10)$, $\lambda=0.80$, $N_{max}=30$.

\

\renewcommand{\arraystretch}{1.3}
\begin{table}[htbp]
\centering
\begin{tabular} {lccclccc} 
\hline
$\bm y$ & $f_{m}({\bm y}|{\bm \alpha}_{E}, {\bm x})$ & $B(\bm Y)$ & $I(B(\bm Y)>\lambda)$ &  $\bm y$ & $f_{m}({\bm y}|{\bm \alpha}_{E}, {\bm x})$ & $B(\bm Y)$ & $I(B(\bm Y)>\lambda)$ \\\hline
$(5,0,0,0)$ & $0.0010619$ & $0.5239$ & $0$ &$(1,2,1,1)$& $0.0102602$ & $0.8384$ & $1$\\
$(4,1,0,0)$ & $0.0058683$ & $0.6415$ & $0$ &$(0,3,1,1)$& $0.0077729$ & $0.9066$ & $1$ \\
$(3,2,0,0)$ & $0.0158789$ & $0.7490$ & $0$ &$(2,0,2,1)$& $0.0004142$ & $0.6393$ & $0$ \\
$(2,3,0,0)$ & $0.0264649$ & $0.8388$ & $1$ &$(1,1,2,1)$& $0.0013383$ & $0.7470$ & $0$ \\
$(1,4,0,0)$ & $0.0274828$ & $0.9069$ & $1$ &$(0,2,2,1)$& $0.0013991$ & $0.8370$ & $1$ \\
$(0,5,0,0)$ & $0.0144909$ & $0.9530$ & $1$ &$(1,0,3,1)$& $0.0001062$ & $0.6365$ & $0$ \\
$(4,0,1,0)$ & $0.0002794$ & $0.5235$ & $0$ &$(0,1,3,1)$& $0.0002028$ & $0.7442$ & $0$ \\
$(3,1,1,0)$ & $0.0013808$ & $0.6412$ & $0$ &$(0,0,4,1)$& $0.0000169$ & $0.6320$ & $0$ \\
$(2,2,1,0)$ & $0.0031758$ & $0.7487$ & $0$ &$(3,0,0,2)$& $0.0158789$ & $0.7489$ & $0$ \\
$(1,3,1,0)$ & $0.0040715$ & $0.8386$ & $1$ &$(2,1,0,2)$& $0.0666916$ & $0.8388$ & $1$ \\
$(0,4,1,0)$ & $0.0024984$ & $0.9067$ & $1$ &$(1,2,0,2)$& $0.1179928$ & $0.9069$ & $1$ \\
$(3,0,2,0)$ & $0.0000986$ & $0.5227$ & $0$ &$(0,3,0,2)$& $0.0893885$ & $0.9530$ & $1$ \\
$(2,1,2,0)$ & $0.0004142$ & $0.6403$ & $0$ &$(2,0,1,2)$& $0.0031758$ & $0.7479$ & $0$ \\
$(1,2,2,0)$ & $0.0007329$ & $0.7479$ & $0$ &$(1,1,1,2)$& $0.0102602$ & $0.8380$ & $1$ \\
$(0,3,2,0)$ & $0.0005552$ & $0.8378$ & $1$ &$(0,2,1,2)$& $0.0107266$ & $0.9063$ & $1$ \\
$(2,0,3,0)$ & $0.0000329$ & $0.5212$ & $0$ &$(1,0,2,2)$& $0.0007329$ & $0.7454$ & $0$ \\
$(1,1,3,0)$ & $0.0001062$ & $0.6387$ & $0$ &$(0,1,2,2)$& $0.0013991$ & $0.8357$ & $1$ \\
$(0,2,3,0)$ & $0.0001110$ & $0.7462$ & $0$ &$(0,0,3,2)$& $0.0001110$ & $0.7408$ & $0$ \\
$(1,0,4,0)$ & $0.0000089$ & $0.5187$ & $0$ &$(2,0,0,3)$& $0.0264649$ & $0.8388$ & $1$ \\
$(0,1,4,0)$ & $0.0000169$ & $0.6359$ & $0$ &$(1,1,0,3)$& $0.0855020$ & $0.9069$ & $1$ \\
$(0,0,5,0)$ & $0.0000014$ & $0.5148$ & $0$ &$(0,2,0,3)$& $0.0893885$ & $0.9529$ & $1$ \\
$(4,0,0,1)$ & $0.0058683$ & $0.6414$ & $0$ &$(1,0,1,3)$& $0.0040715$ & $0.8373$ & $1$ \\
$(3,1,0,1)$ & $0.0289963$ & $0.7489$ & $0$ &$(0,1,1,3)$& $0.0077729$ & $0.9058$ & $1$ \\
$(2,2,0,1)$ & $0.0666916$ & $0.8388$ & $1$ &$(0,0,2,3)$& $0.0005552$ & $0.8334$ & $1$ \\
$(1,3,0,1)$ & $0.0855020$ & $0.9069$ & $1$ &$(1,0,0,4)$& $0.0274828$ & $0.9068$ & $1$ \\
$(0,4,0,1)$ & $0.0524672$ & $0.9530$ & $1$ &$(0,1,0,4)$& $0.0524672$ & $0.9529$ & $1$ \\
$(3,0,1,1)$ & $0.0013808$ & $0.6408$ & $0$ &$(0,0,1,4)$& $0.0024984$ & $0.9048$ & $1$ \\
$(2,1,1,1)$ & $0.0057993$ & $0.7484$ & $0$ &$(0,0,0,5)$& $0.0144909$ & $0.9529$ & $1$ \\\hline

\end{tabular}
\end{table}

\newpage
\hypertarget{T5}{}
\noindent
{\bf{Table 5:}}
Description of the scenarios of interest and the probability of early termination (PET), the percentage of rejecting the null hypothesis (PRN), and the average sample size (ASS) when ${\bm \alpha}_{E}=(0.5, 0.5, 0.5, 0.5)$, ${\bm \alpha}_{S}=(30, 60, 30, 80)$, $\lambda=0.80$, $\theta_{L}=0.001$, $N_{min}=10$, $N_{max}=40$ with 10,000 simulated trials.

\

\renewcommand{\arraystretch}{1.3}
\begin{table}[htbp]
\centering
\begin{tabular} {lccccccc} 
\hline
&& \multicolumn{3}{c}{Cohort size: 1 participant} &\multicolumn{3}{c}{Cohort size: 5 participants}\\ \cline{3-5} \cline{6-8}
Scenario   & ${\bm p}_{E}$        &   ~    PET  ~   & ~ PRN ~ &  ~ASS~ &      ~ PET ~    & ~ PRN ~ & ~ ASS  ~\\\hline
1.   & $(0.15, 0.30, 0.15, 0.40)$     &  0.9292  & 0.0550 & 25.0649 & 0.8287 & 0.0531 & 27.3005 \\
2.   & $(0.15, 0.30, 0.20, 0.35)$     &  0.9793  & 0.0146 & 20.9999 & 0.9403 & 0.0139 & 23.0835 \\
3.   & $(0.20, 0.25, 0.15, 0.40)$     &  0.9815  & 0.0121 & 20.9437 & 0.9364 & 0.0135 & 23.1800 \\
4.   & $(0.20, 0.20, 0.10, 0.50)$     &  0.9422  & 0.0424 & 24.3711 & 0.8490 & 0.0407 & 26.5900 \\
5.   & $(0.15, 0.15, 0.15, 0.55)$     &  0.9864  & 0.0087 & 20.0937 & 0.9440 & 0.0082 & 22.4240 \\
6.   & $(0.18, 0.42, 0.02, 0.38)$     &  0.3810 & 0.5863 & 36.6277 & 0.2618 & 0.5858 & 37.3565 \\
7.   & $(0.18, 0.46, 0.02, 0.34)$     &  0.3758 & 0.5937 & 36.7275 & 0.2540 & 0.5938 & 37.4615 \\
8.   & $(0.15, 0.45, 0.02, 0.38)$     &  0.2124 & 0.7645 & 38.3488 & 0.1328 & 0.7695 & 38.7730 \\
9.   & $(0.10, 0.50, 0.05, 0.35)$     &  0.1327 & 0.8486 & 39.0428 & 0.0745 & 0.8533 & 39.3100\\
10.  & $(0.10, 0.55, 0.05, 0.30)$    &  0.1246 & 0.8592 & 39.1270 & 0.0693 & 0.8649 & 39.3545 \\\hline
\end{tabular}
\end{table}

\

\

\hypertarget{T6}{}
\noindent
{\bf{Table 6:}}
Description of the scenarios of interest assumed with various true no efficacy and toxicity and the probability of early termination (PET), the percentage of rejecting the null hypothesis (PRN), and the average sample size (ASS) when ${\bm \alpha}_{E}=(0.5, 0.5, 0.5, 0.5)$, ${\bm \alpha}_{S}=(30, 60, 30, 80)$, $\lambda=0.80$, $\theta_{L}=0.001$, $N_{min}=10$, $N_{max}=40$ with 10,000 simulated trials. 

\

\renewcommand{\arraystretch}{1.3}
\begin{table}[htbp]
\centering
\begin{tabular} {lccccccc} 
\hline
&& \multicolumn{3}{c}{Cohort size: 1 participant} &\multicolumn{3}{c}{Cohort size: 5 participants}\\ \cline{3-5} \cline{6-8}
Scenario   & ${\bm p}_{E}$        &   ~    PET  ~   & ~ PRN ~ &  ~ASS~ &      ~ PET ~    & ~ PRN ~ & ~ ASS  ~\\\hline
1.   & $(0.15, 0.30, 0.20, 0.35)$    &  0.9813  & 0.0145 & 20.8878 & 0.9408 & 0.0134 & 22.9910 \\
2.   & $(0.19, 0.30, 0.16, 0.35)$    &  0.9764  & 0.0181 & 21.6474 & 0.9305 & 0.0178 & 23.6330 \\
3.   & $(0.23, 0.30, 0.12, 0.35)$    &  0.9707  & 0.0217 & 22.2601 & 0.9189 & 0.0232 & 24.2165 \\
4.   & $(0.27, 0.30, 0.08, 0.35)$    &  0.9667  & 0.0264 & 22.8911 & 0.9223 & 0.0261 & 24.6960 \\
5.   & $(0.31, 0.30, 0.04, 0.35)$    &  0.9570  & 0.0325 & 23.1372 & 0.9121 & 0.0286 & 24.8230 \\\hline
\end{tabular}
\end{table}

\newpage
\hypertarget{T7}{}
\noindent
{\bf{Table 7:}}
Type I error and power values under the null and alternative hypotheses by varying the design parameter $(\lambda, \theta_{L})$ when $\bm{P}_{E}=(0.15, 0.30, 0.15, 0.40)$ under $H_{0}$, $\bm{P}_{E}=(0.10, 0.50, 0.05, 0.35)$ under $H_{1}$, ${\bm \alpha}_{E}=(0.5, 0.5, 0.5, 0.5)$, ${\bm \alpha}_{S}=(30, 60, 30, 80)$, $N_{min}=10$, $N_{max}=40$, and the cohort size equals to five participants with 10,000 simulated trials.

\

\renewcommand{\arraystretch}{1.3}
\begin{table}[htbp]
\centering
\begin{tabular} {lccccccccccc} 
\hline
$\theta_{L}$ & \multicolumn{5}{c}{Null hypothesis with $\lambda$} & & \multicolumn{5}{c}{Alternative hypothesis with $\lambda$}    \\ \cline{2-6} \cline{8-12}
       & 0.70      & 0.75 & 0.80 & 0.85 & 0.90 &  & 0.70 & 0.75 & 0.80 & 0.85 & 0.90\\\hline
0.001 & \multicolumn{1}{c|}{0.1146} & 0.0852 & 0.0578 & 0.0355 & 0.0179 && 0.9238 & 0.9137 & 0.8558 & \multicolumn{1}{c|}{0.8226} & 0.7115 \\
	0.01  & \multicolumn{1}{c|}{0.1099} &  \cellcolor[gray]{0.8}0.0813 & 0.0508 & 0.0355 & 0.0170 && 0.9212 & \cellcolor[gray]{0.8}0.9150 & 0.8484 & \multicolumn{1}{c|}{0.8100} & 0.7145 \\ \cline{2-2} \cline{11-11}
0.05  & 0.0967 & 0.0760 & 0.0492 & 0.0321 & 0.0169 && 0.8970 & 0.8902 & \multicolumn{1}{c|}{0.8227} & 0.7806 & 0.6775 \\\cline{10-10}
0.10  & 0.0909 & 0.0717 & 0.0451 & 0.0285 & 0.0139 && 0.8733 & \multicolumn{1}{c|}{0.8659} & 0.7827 & 0.7501 & 0.6393 \\\hline
\end{tabular}
\end{table}

\newpage
\hypertarget{F1}{}
\noindent
{\bf{Figure 1:}}
Posterior distribution using Monte Carlo approach with 10,000 times repeat simulations when ${\bm \alpha}_{E}=(0.5, 0.5, 0.5, 0.5)$, ${\bm \alpha}_{S}=(10, 9, 11, 30)$, ${\bm x}=(5, 10, 0, 10)$, ${\bm y}=(0, 3, 2, 0)$, $N_{max}=30$. (a) presents ${\bm \Phi}_{t}$ for $t=\rm{E, S}$ and (b) presents the difference distribution ${\bm \Phi}_{E}-{\bm \Phi}_{S}$. E and S represent experimental and standard treatments, respectively.

\

\begin{figure}[h]
\centering
\includegraphics[scale=0.5]{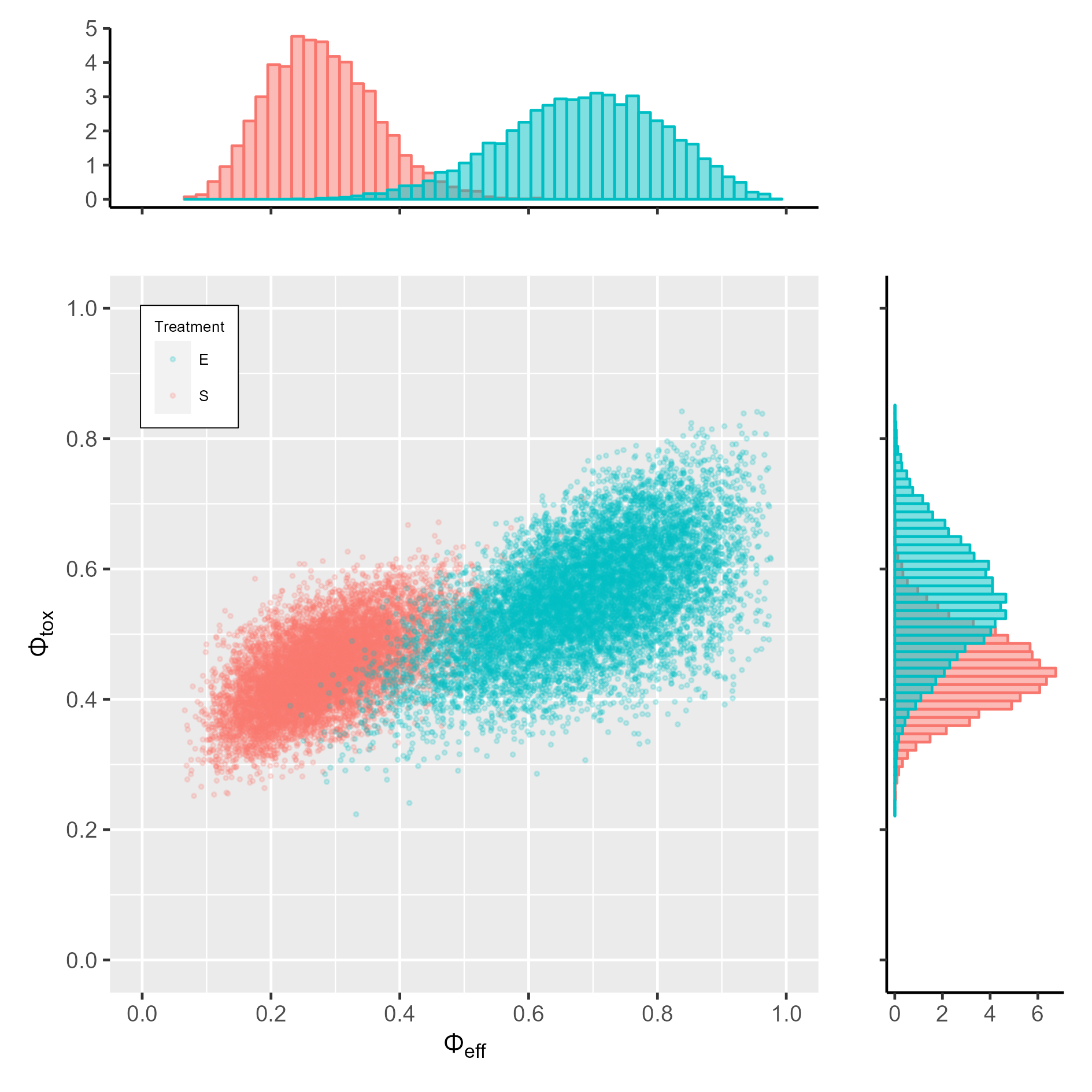}
\captionsetup{labelformat=empty,labelsep=none}
\caption{(a) Simulation distribution of ${\bm \Phi}_{E}$ and ${\bm \Phi}_{S}$}
\end{figure}

\

\begin{figure}[h]
\centering
\includegraphics[scale=0.5]{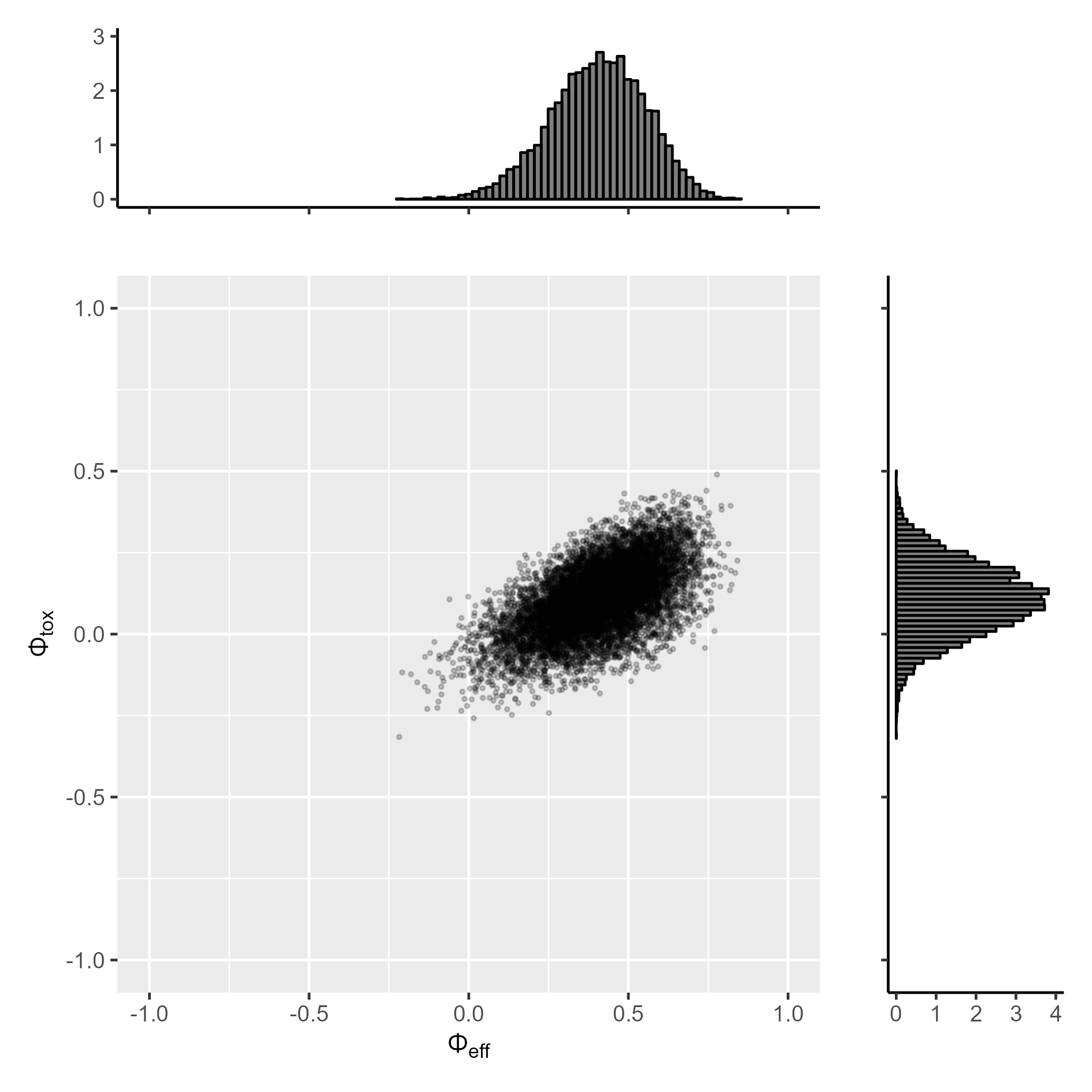}
\captionsetup{labelformat=empty,labelsep=none}
\caption{(b) Simulation distribution of ${\bm \Phi}_{E}-{\bm \Phi}_{S}$}
\end{figure}

\newpage
\hypertarget{F2}{}
\noindent
{\bf{Figure 2:}}
Posterior distribution using asymptotic approach when ${\bm \alpha}_{E}=(0.5, 0.5, 0.5, 0.5)$, ${\bm \alpha}_{S}=(10, 9, 11, 30)$, ${\bm x}=(5, 10, 0, 10)$, ${\bm y}=(0, 3, 2, 0)$,  $N_{max}=30$. (a) illustrates ${\bm \Phi}_{t}$ for $t=\rm{E, S}$ and (b) illustrates the difference distribution ${\bm \Phi}_{E}-{\bm \Phi}_{S}$. E and S represent experimental and standard treatments, respectively.

\

\begin{figure}[h]
\centering
\includegraphics[scale=0.5]{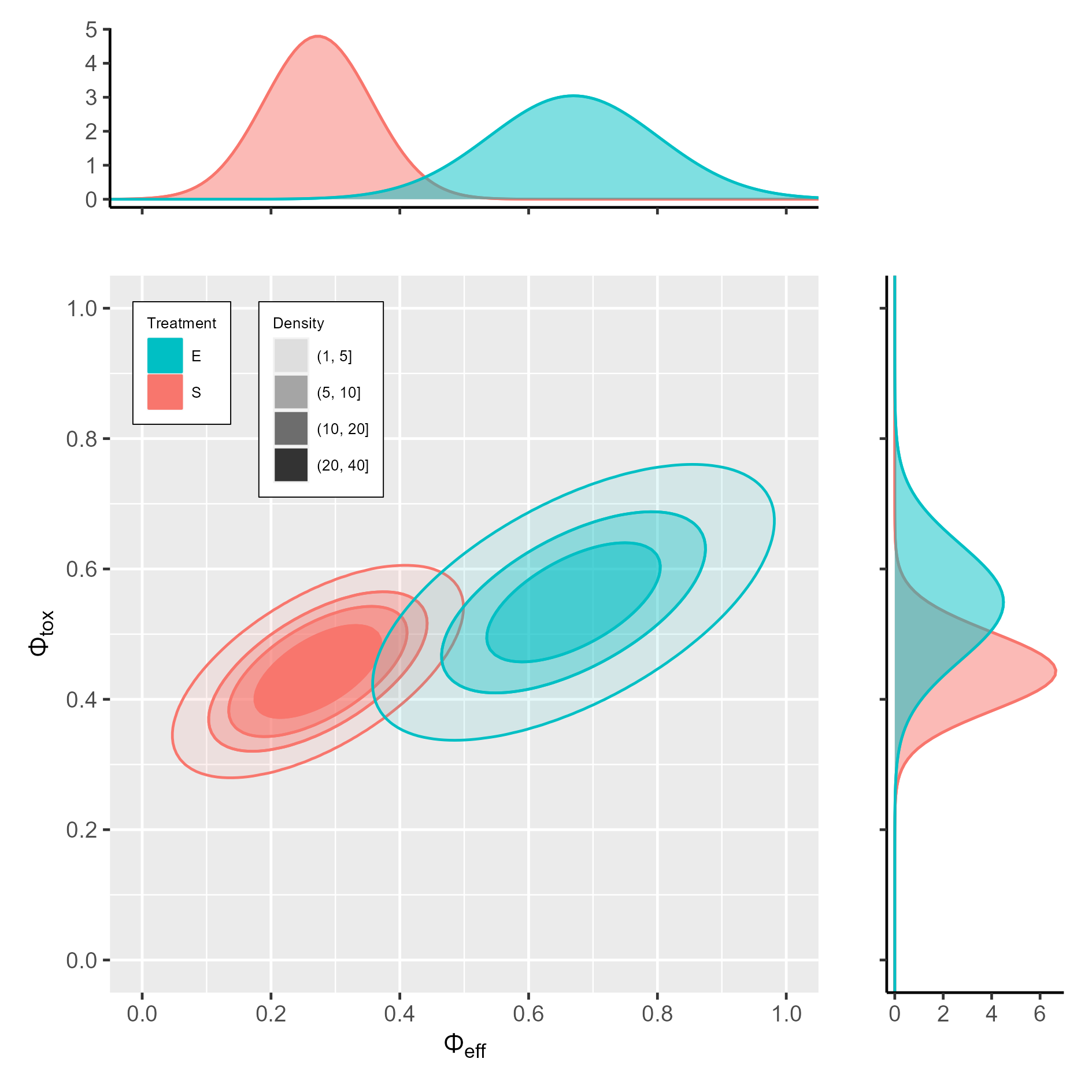}
\captionsetup{labelformat=empty,labelsep=none}
\caption{(a) Asymptotic distribution of ${\bm \Phi}_{E}$ and ${\bm \Phi}_{S}$}
\end{figure}

\

\begin{figure}[h]
\centering
\includegraphics[scale=0.5]{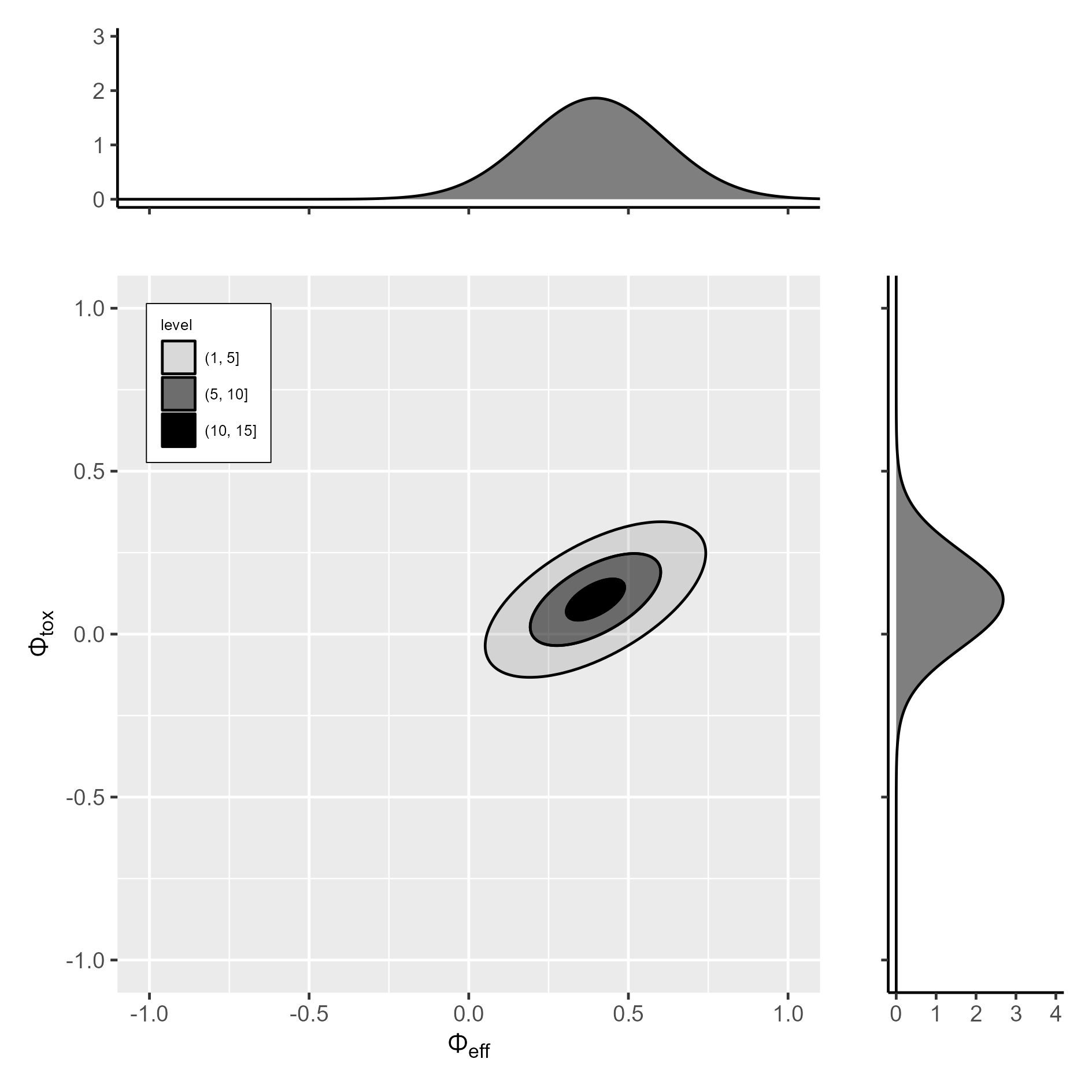}
\captionsetup{labelformat=empty,labelsep=none}
\caption{(b) Asymptotic distribution of ${\bm \Phi}_{E}-{\bm \Phi}_{S}$}
\end{figure}

\newpage
\hypertarget{F3}{}
\noindent
{\bf{Figure 3:}}
Predictive probability under various settings when $\bm{P}_{E}=(0.15, 0.30, 0.15, 0.40)$ under $H_{0}$, $\bm{P}_{E}=(0.10, 0.50, 0.05, 0.35)$ under $H_{1}$, ${\bm \alpha}_{E}=(0.5, 0.5, 0.5, 0.5)$, $\lambda=0.80$, and $m=20$. (a) and (b) illustrate the predictive probability under $H_{0}$ and $H_{1}$, respectively. E and S represent experimental and standard treatments, respectively.

\

\begin{figure}[h]
\centering
\includegraphics[scale=0.5]{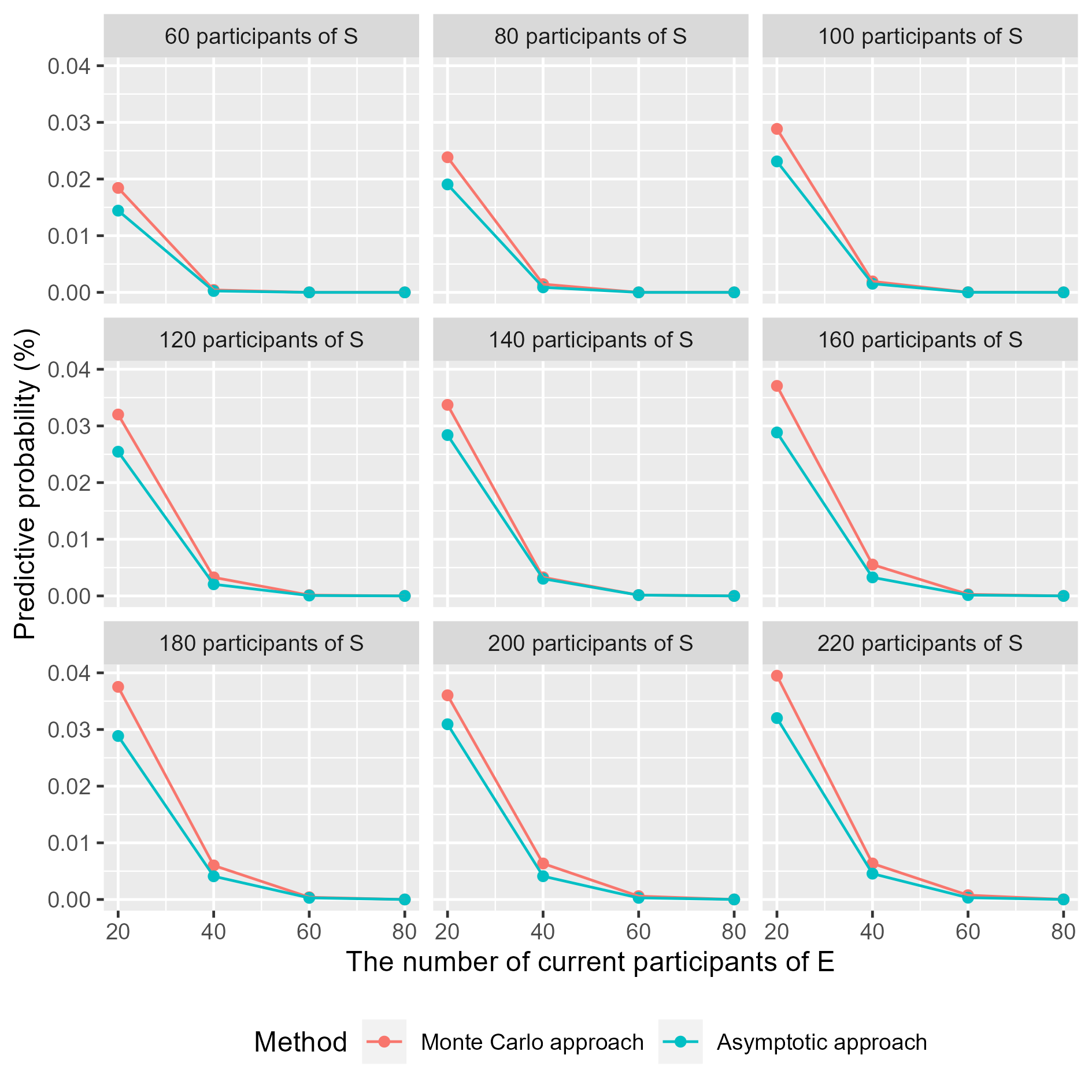}
\captionsetup{labelformat=empty,labelsep=none}
\caption{(a) Predictive probability under $H_{0}$}
\end{figure}

\

\begin{figure}[h]
\centering
\includegraphics[scale=0.5]{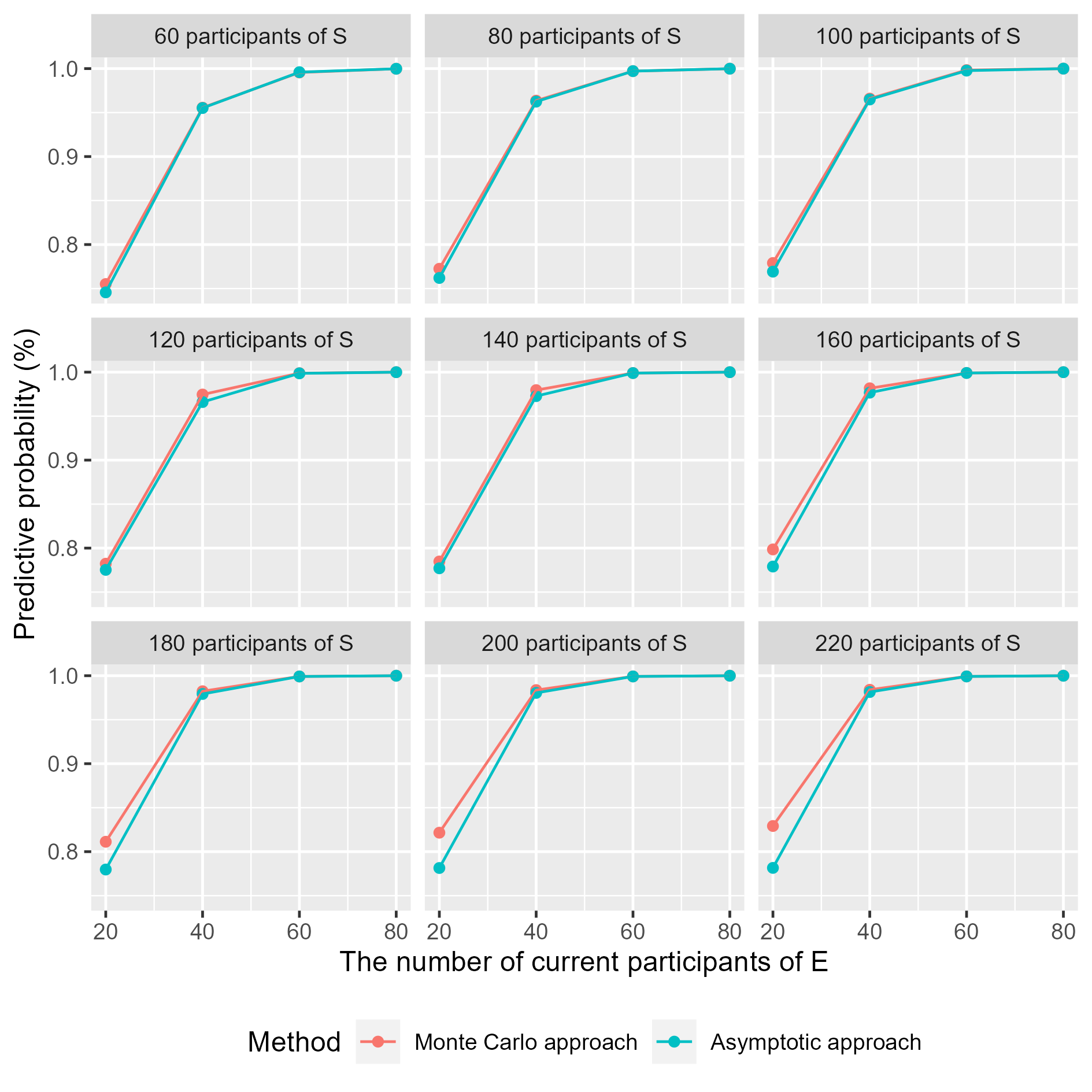}
\captionsetup{labelformat=empty,labelsep=none}
\caption{(b) Predictive probability under $H_{1}$}
\end{figure}

\end{document}